\documentclass[aps,preprintnumbers,showpacs,twocolumn,superscriptaddress,floatfix,nofootinbib]{revtex4-1}

\usepackage{latexsym} 
\usepackage{amssymb} 
\usepackage{amsmath} 
\usepackage{amsfonts}
\usepackage{bm}
\usepackage{xcolor}
\usepackage{times} 
\usepackage{units}
\usepackage{hyperref}
\usepackage[normalem]{ulem}
\usepackage[utf8x]{inputenc}
\usepackage{graphicx} 
\usepackage[squaren]{SIunits} 
\usepackage[inline]{enumitem}

\newcommand{\Eref}[1]{Eq.~\eqref{#1}} 
\newcommand{\Fref}[1]{Fig.~\ref{#1}} 
\newcommand{\Sref}[1]{Sec.~\ref{#1}}

\begin{document}

%=====================================================================
% Header
%=====================================================================

\title{Implementing a new recovery scheme for primitive variables in the general relativistic magnetohydrodynamic code Spritz}

\author{Jay V. Kalinani}
\address{Universit\`a di Padova, Dipartimento di Fisica e Astronomia, Via Francesco Marzolo 8, I-35131 Padova, Italy}
\address{INFN, Sezione di Padova, Via Francesco Marzolo 8, I-35131 Padova, Italy}

\author{Riccardo Ciolfi}
\address{INAF, Osservatorio Astronomico di Padova, Vicolo dell'Osservatorio 5, I-35122 Padova, Italy}
\address{INFN, Sezione di Padova, Via Francesco Marzolo 8, I-35131 Padova, Italy}

\author{Wolfgang Kastaun}
\address{Max Planck Institute for Gravitational Physics (Albert
  Einstein Institute), Callinstrasse 38, 30167 Hannover, Germany}
\address{Leibniz Universit\"at Hannover, 30167 Hannover, Germany}

\author{Bruno Giacomazzo}
\address{Universit\`a degli Studi di Milano - Bicocca, Dipartimento di Fisica G. Occhialini, Piazza della Scienza 3, I-20126 Milano, Italy}
\address{INFN, Sezione di Milano-Bicocca, Piazza della Scienza 3, I-20126 Milano, Italy}
\address{INAF, Osservatorio Astronomico di Brera, Via E. Bianchi 46, I-23807 Merate (LC), Italy}

\author{Federico Cipolletta}
\address{Leonardo Corporate LABS, Via Raffaele Pieragostini 80, 16149 Genova (GE), Italy}
\address{Center for Computational Relativity and Gravitation, School of Mathematical Sciences, Rochester Institute of Technology, 85 Lomb Memorial Drive, Rochester, New York 14623, USA}

\author{Lorenzo Ennoggi}
\address{Center for Computational Relativity and Gravitation, School of Mathematical Sciences, Rochester Institute of Technology, 85 Lomb Memorial Drive, Rochester, New York 14623, USA}

%=====================================================================
% Abstract
%=====================================================================
\begin{abstract}
\noindent 
General relativistic magnetohydrodynamic (GRMHD) simulations represent a fundamental tool to probe various underlying mechanisms at play during binary neutron star (BNS) and neutron star (NS) - black hole (BH) mergers. Contemporary flux-conservative GRMHD codes numerically evolve a set of conservative equations based on `conserved' variables which then need to be converted back into the fundamental (`primitive') variables. The corresponding conservative-to-primitive variable recovery procedure, based on root-finding algorithms, constitutes one of the core elements of such GRMHD codes. Recently, a new robust, accurate and efficient recovery scheme called RePrimAnd was introduced, which has demonstrated the ability to always converge to a unique solution. The scheme provides fine-grained error policies to handle invalid states caused by evolution errors, and also provides analytical bounds for the error of all primitive variables.  In this work, we describe the technical aspects of implementing the RePrimAnd scheme into the GRMHD code Spritz. To check our implementation as well as to assess the various features of the scheme, we perform a number of GRMHD tests in three dimensions. Our tests, which include critical cases such as a NS collapse to a BH as well as the early evolution ($\sim\!50$\,ms) of a Fishbone-Moncrief BH-accrection disk system, show that RePrimAnd is able to support magnetized, low density environments with magnetic-to-fluid pressure ratios as high as $10^4$, in situations where the previously used recovery scheme fails.
\end{abstract}

\maketitle

%=====================================================================
%=====================================================================
\section{Introduction}
\label{sec:intro}

\noindent At the advent of multi-messenger astronomy with gravitational wave (GW) sources, compact binary coalescences involving at least one neutron star (NS) present a rich phenomenology with the potential to answer key open questions in astrophysics and fundamental physics. 
The coincident detection of GWs from the binary neutron star (BNS) merger GW170817 and the accompanying electromagnetic counterparts including a short gamma-ray burst (SGRB) and an optical/infrared kilonova (named GRB\,170817A and AT2017gfo, respectively) offers a striking example \cite{LVC-BNS,LVC-GRB,LVC-MMA}.
This single event has cemented the connection between BNS mergers, SGRBs, and radioactively powered kilonovae and, at the same time, has led to the first GW-based constraints on the NS equation of state (EOS) and the Hubble constant 
(e.g., \cite{Lazzati2018,Mooley2018b,Ghirlanda2019,Kasen2017,Pian2017,LVC-170817properties,LVC-Hubble,LVC-GRB,LVC:EOSModelSel:2020}; see also, e.g., \cite{Ciolfi2020c} and refs.~therein).

General relativistic magnetohydrodynamic (GRMHD) simulations provide the essential framework necessary to investigate various astrophysical processes involved in BNS and NS-black hole (BH) mergers, including magnetic field effects as a fundamental element (e.g., \cite{Ciolfi2020b,Ruiz2021} and refs.~therein). In the past decade, various numerical relativity groups have successfully performed stable BNS or NS-BH merger simulations in GRMHD, covering a growing range of the parameter space and adding more and more key physical ingredients such as temperature and composition dependent tabulated EOS, neutrino radiation, and/or NS spin (e.g., \cite{Palenzuela2015,Paschalidis2015,Kawamura2016,Ciolfi2017,Most2019,Ruiz2019,Moesta2020,Ruiz2020,East2021,Most2021} and refs.~therein).

Recently, we introduced a new publicly available GRMHD code called {\tt Spritz} \footnote{The latest version of {\tt Spritz} is publicly available on Zenodo \cite{Giacomazzo2020}.} \cite{Cipolletta2020}, built to work within the {\tt Einstein Toolkit} infrastructure \cite{EinsteinToolkit1,EinsteinToolkit2, EinsteinToolkit3}. We further improved and tested the code as described in \cite{Cipolletta2021}. It now includes support for high-order numerical schemes, use of finite temperature tabulated EOS, as well as a ray-by-ray neutrino leakage scheme that is based on the {\tt ZelmaniLeak} public code and which accounts for both neutrino cooling and heating \cite{OConnor2010,Ott2013}.

In GRMHD codes like {\tt Spritz}, the implementation of GRMHD equations is based on a flux-conservative formalism which evolves a set of ``conserved'' variables that need to be converted back to the ``primitive'' variables. The corresponding conservative-to-primitive (C2P) recovery procedure represents a central constituent of any GRMHD code. Since there is no generic analytical solution to the problem, codes resort to numerical root-finding procedures. However, finding a robust formulation of the problem is not straightforward. The C2P scheme is often an error-prone component of GRMHD codes that tends to fail in certain problematic regimes. 
A recent study \cite{Siegel2018c2p} subjected available state of the art C2P schemes to a battery of test cases.
It was found that the various schemes started failing at Lorentz factors 10--1000 even for zero magnetization,
and that increasing magnetization tends to deteriorate the robustness further (see Fig. 3 in \cite{Siegel2018c2p}). 
Moreover, schemes based on Newton-Raphson root solvers require an initial guess, and, as pointed out in 
\cite{Siegel2018c2p}, the robustness depends strongly on the accuracy of the initial guess.
Since evolution schemes rely on previous timesteps for the initial guess, the robustness suffers further in highly
dynamic regimes. The astrophysically important regime of low density regions within strong magnetic fields that
occurs near merger remnants is therefore very problematic. 
In practice, evolution schemes artificially enforce a minimum density (since the evolution equations degenerate in
vacuum) and the robustness of the C2P can be a limiting factor for this choice.
Another frequent cause for recovery failures is when evolution errors result in unphysical states of the evolved variables. 

In order to overcome such limitations, we recently designed a new C2P scheme \cite{Kastaun2021} and made a reference implementation publicly available \cite{Kastaun2020Reprimand} in form of a stand-alone library named \texttt{RePrimAnd}. This library offers a novel C2P algorithm  (called RePrimAnd or RPA in the following) based on a formulation 
for which existence and uniqueness of the solution was proven mathematically.
As a standalone version, the scheme has undergone rigorous testing and proven to be robust, accurate and efficient even in extremely problematic regimes \cite{Kastaun2021}. 
The \texttt{RePrimAnd} library was already used successfully in a few BNS merger simulations \cite{Aguilera2020}.

In this work, we incorporate the C2P implementation provided by the \texttt{RePrimAnd} library into the {\tt Spritz} GRMHD code. Our primary goal is to test the interplay between the RePrimAnd C2P scheme and the evolution code, as well as to further enhance the capabilities of the {\tt Spritz} code. Therefore, we subject the code to a set of increasingly demanding three dimensional test problems. 
As a by-product, the necessary simulations can also serve as additional tests of the 
evolution code itself, although our focus remains on the new C2P implementation.
We present a direct comparison to the widely-used C2P scheme of \cite{Noble2006}, which is based on a 2D Newton-Raphson root-finding technique.
This scheme is already implemented in the public {\tt Spritz} code \cite{Cipolletta2020}. Throughout this article, we will refer to
this implementation as `the Noble scheme'.

The use of the RePrimAnd scheme also enables the code to reliably recognize invalid states of the evolved variables, and to apply corrections for errors deemed harmless according to an explicit error-policy. We will discuss such policies employed in our tests. 

The paper is organized as follows. In \Sref{sec:formulation}, a short summary of the RePrimAnd scheme along with the steps taken to implement it into {\tt Spritz} is provided. We describe the numerical setup and the results of our different 3D tests in \Sref{sec:tests}. 
Finally, we present our conclusions and outlook in \Sref{sec:conclusions}. 
In Appendix \ref{STtests}, we summarize additional 1D shocktube tests.
Unless noted otherwise, we use geometric units $G=c=M_\odot=1$.
We use two different dimensionless quantities to express the degree of magnetization. The ratio of magnetic pressure to fluid pressure in the comoving frame is denoted by $\beta^{-1}$, whereas $b$ denotes the ratio $B/\sqrt{W\rho}$, with Lorentz factor $W$, baryonic mass density $\rho$, and magnetic field strength in the Eulerian frame $B$.

%=====================================================================
%=====================================================================
\section{Recovery scheme implementation}
\label{sec:formulation}

\noindent In this section, we describe the aspects of the RePrimAnd scheme that are relevant for the integration into the {\tt Spritz} code. 
For details of the recovery algorithm, we refer to \cite{Kastaun2021} (see Fig.~2 therein for 
a schematic overview).

The scheme yields the primitive variables defined as follows. In the fluid frame,
$\rho$ denotes the  baryonic mass density,
$Y_e$ the electron fraction,
$\rho_E$ the fluid contribution to the total energy density,
$\epsilon = \frac{\rho_E}{\rho} - 1$ the specific internal energy,
$h = 1 + \epsilon + \frac{P}{\rho}$ the relativistic enthalpy,
and $P$ the fluid pressure. 
In the Eulerian frame, 
$W$ is the fluids' Lorentz factor,
$v^i$ its 3-velocity,
$B^i$ the magnetic field,
and $E^i$ the electric field. 
The magnetic field is defined
\footnote{Note there is also a competing convention for the magnetic field that differs by a factor $\sqrt{4\pi}$.}
as
$B^\mu  = n_\nu {}^*F^{\mu\nu}$ 
in terms of the Faraday tensor $F^{\mu\nu}$, where the star denotes the dual and
$n_\nu$ is the 4-velocity of the Eulerian observer.

As input the scheme expects the following ``conserved'' variables:
\begin{align}
\label{cv}
D &=\sqrt{\gamma} \left[ \rho W \right], \\
\tau &= \sqrt{\gamma} \left[ \rho W(hW-1) - P + \frac{1}{2}(E^2 + B^2) \right], \\
S_i &= \sqrt{\gamma} \left[ \rho W^2 h v_i + \epsilon_{ijk}E^j B^k \right], \\
B^i_c &= \sqrt{\gamma} \left[ B^i \right],\\
Y_E^c  &= D Y_E, 
\end{align}
where $\sqrt{\gamma}$ is the volume element of the 3-metric $g_{ij}$.
Those variables are readily available in {\tt Spritz}. $D, \tau, S_i$ are evolved 
variables. $B^i$ is not evolved, but is computed directly from the evolved vector 
potential $A_\mu$. The ``tracer'' variable $Y_E^c$ is evolved using a conservation-law
formulation corresponding to advection of the electron fraction along fluid trajectories,
with additional source terms in case neutrino transport is activated.

The RePrimAnd scheme is designed exclusively around the assumption of ideal MHD,
which implies that the electric field is computed using
\begin{equation}
\label{efield}
E^i = -\epsilon^{ijk}v_j B_k .
\end{equation}
C2P schemes for ideal MHD are not compatible with evolutionary GRMHD codes which work beyond this simplified assumption. 
To include, for instance, the effects of resistivity, fundamentally changes the C2P problem and requires
completely different C2P schemes. For further discussion, we refer to \cite{Palenzuela2009,Dionysopoulou2013,Ripperda2019,Wright2020}.

The scheme further requires an EOS of the form $P(\rho, \epsilon, Y_e)$, where EOS that do not 
consider the electron fraction are emulated by treating it as a dummy variable. 
Derivatives of the EOS are not used in any way by the RePrimAnd recovery scheme. It does
however require a well defined validity region for the EOS in order to judge if 
a given combination of conserved variables is physically valid.

In short, the RePrimAnd scheme works by casting the recovery problem into a one-dimensional 
root-finding problem, choosing the independent variable as $\mu = (Wh)^{-1}$. 
The root is determined using a root-bracketing scheme that does not require derivatives. 
The scheme includes a prescription for the initial bracket. 
There is a mathematical proof that the root is unique and contained in the initial bracket \cite{Kastaun2021}. 
Consequently, the RePrimAnd scheme does not 
require an initial guess, in contrast to other schemes such as the Noble scheme.

An important property of the system of nonlinear equations for computing the conserved from the primitive 
variables is that it can not always be inverted. There are combinations of conserved variables
that do not correspond to any physically valid primitive variables. The RePrimAnd scheme
is able to distinguish reliably between valid and invalid input, and even provides more 
fine-grained information about which constraints on the primitives are violated.

The RePrimAnd scheme offers various corrections that map invalid input back onto 
valid input (see \cite{Kastaun2021} for a complete list). These are optional and governed 
by a user-provided error policy. For any corrections, conservative and primitive
variables are always kept consistent, recomputing the conservatives if necessary.
It should be pointed out that the error policy and 
corrections are completely agnostic regarding the type of EOS. 
{\tt Spritz} makes use of those optional corrections when using the RePrimAnd scheme.

The most frequent but harmless type of invalid input is given by specific internal energy
that would have to be below the one for zero temperature. This is bound to happen 
when evolving zero-temperature initial data. The policy we adopt for this case 
is to change the evolved energy to the zero-temperature value. 

Another correction concerns the NS surface. When the surface is moving, e.g. 
during NS oscillations, the density in a numerical cell at the surface can change 
drastically during a single timestep, leading in rare cases to extreme velocities
and/or internal specific energy at densities near the atmosphere cutoff.
To alleviate this problem, we employ a policy that limits the velocity by means of 
rescaling the momentum and also limits the specific internal energy (in this work we 
use limits $W\le100$ and $\epsilon\le 50$). Since the problem should only occur 
near the surface of NSs but not in the interior, our policy is to treat larger values
as an error if the density is above a threshold $\rho_{\rm strict}$ (in this work
$\rho_{\rm strict}=3.95\times 10^{13}$\,g/cm$^3$, a few percent of the maximum density).
We note that the problems at the NS surface are not specific to MHD but occur already 
for purely hydrodynamic simulations. For such simulations a similar error policy
has been used in several earlier works, e.g. \cite{Galeazzi:2013:64009,Kastaun:2015:064027,Kastaun:2016}.
We emphasize that the above corrections are not required to prevent C2P 
failures, in particular the velocity rescaling is employed \emph{after} the C2P solution is found.

RePrimAnd allows to specify hard limits for the magnetic scale variable 
$b=B / \sqrt{W\rho}$, which are treated as error when exceeded (we stress 
that the magnetic field is never adjusted in any way). This merely serves as a safeguard 
to catch potential evolution errors that might otherwise go unnoticed, given the robustness
of the C2P scheme. In our tests, we set this limit to $b_{\rm max}=100.0$, since such 
a large value would be unexpected. 
The limit was indeed never reached in any of our tests.

Another optional correction is to enforce the electron fraction to
stay within the EOS validity range. This option is irrelevant for our tests since
we do not include neutrino reactions and use EOS for which $Y_e$ is an unused 
dummy parameter.

When using the Noble scheme in {\tt Spritz} instead of the RePrimAnd scheme, the above 
corrections are not applied. 
For example, the formulas defining the hybrid gamma-law or ideal gas
EOS are applied to any value of specific energy when using the Noble scheme,
whereas the EOS interface from the \texttt{RePrimAnd} library which is used in the new C2P 
enforces strict physical bounds even for analytic EOS.
This implies that our comparisons between the two cases
do not just compare the C2P scheme, but also the error policies. We will come back to 
this point.

In terms of the absolute error,
it might even be advantageous to allow slightly negative temperatures. As discussed
in \cite{Kastaun2021}, the error distribution might be skewed by mapping negative 
temperatures to zero, causing more heating on average. However, applying different corrections for every type of EOS can jeopardize the reliability of the tests.

A special case for corrections concerns the interior of BHs. Since the center is always severely underresolved, the numerical discretization errors can become much larger, but at the same time, their impact is limited by the presence of the horizon.
When using the RePrimAnd scheme, {\tt Spritz} is therefore employing a different, more lenient error policy near the center of BHs. If the lapse function is below a given threshold, we enforce the aforementioned speed limit at any density (and 
also allow effectively unlimited values for $b$, although we have no indication that is necessary).

An important point concerns the artificial atmosphere that is required because the evolution scheme cannot handle
vacuum. We employ the standard method of forcing the density to stay above a floor density $\rho_\text{atmo}$,
and set the fluid velocity to zero when the density became smaller. Further, we chose to set the temperature
to zero in the atmosphere. We note that a zero-velocity artificial atmosphere 
also has consequences for the magnetic field evolution because of the ideal MHD condition Eq.~(\ref{efield}).
Taking into account the evolution equations used in \texttt{Spritz}, one can see that
both vector potential and magnetic fields remain frozen in regions covered by the atmosphere.
The atmosphere is enforced by the C2P schemes in essentially the same way.

Another useful feature of the RePrimAnd recovery scheme is the availability of errors bounds 
for all primitive variables based on a single accuracy parameter $\Delta$, which can be 
prescribed by the user. The accuracy of individual primitive variables is derived by means 
of error propagation, with explicit formulas Eqs.~(69)-(73) given in \cite{Kastaun2021}.
Of course, the above does not cover evolution errors.
In our tests, we adopt a value $\Delta=10^{-8}$.

The scheme was subjected to exhaustive stand-alone tests in \cite{Kastaun2021}, exploring a 
parameter space expected to cover all scenarios arising in BNS merger simulations.
The variables most relevant for the shape of the root-solving master function are
the Lorentz factor $W$ and a scale variable $b^2= B^2 / (W\rho)$ for the magnetic field 
strength. The magnetic field becomes irrelevant for the root finding if $\mu b^2 \ll 1$.
The tests in \cite{Kastaun2021} covered the full parameter space within limits
$W< 1000, b<5, \epsilon < 50$, without any errors. 
In terms of performance, the tests suggested a low average number of EOS calls ($<10$)
for the parameter space of typical BNS simulations. We therefore expect a good 
performance of the algorithm. The computational costs in the context of 
our test simulations will be quantified in Sec.~\ref{sec:magtov}.

The RePrimAnd scheme is implemented as a stand-alone C++ library called {\tt RePrimAnd} \cite{Kastaun2020Reprimand}. 
In order to employ the scheme with {\tt Spritz}, we created a module for the {\tt Einstein Toolkit} that integrates the {\tt RePrimAnd} library, and added code to {\tt Spritz} that makes use of the {\tt RePrimAnd} C2P functionality.

The library also includes an EOS framework that provides a fully generic interface to different types of EOS. 
It currently implements the classical ideal gas EOS and the hybrid EOS (which combines an arbitrary cold EOS with a gamma-law thermal component). 
The extension to fully tabulated EOS mainly entails the handling of quality issues often encountered with nuclear physics tables and will be the topic of another work.
The new EOS framework is only used by the new recovery algorithm, while {\tt Spritz} itself still uses  
an EOS framework provided by {\tt Einstein Toolkit}. One reason for this is that the evolution 
code does not take into account any EOS validity bounds during the reconstruction step. 
The new EOS framework strictly forbids invalid arguments, however. 
Making the reconstruction code aware of validity ranges would be a development task 
involving design decisions and tests completely unrelated to the primitive recovery we are 
studying in this work.

%=====================================================================
%=====================================================================
\section{Tests}
\label{sec:tests}

\noindent In this section, we study the interplay between the novel C2P scheme and the numerical time evolution 
scheme, and ascertain the correct implementation of the interface between evolution code and the \texttt{RePrimAnd} library. 
To this end, we perform a series of demanding 3D tests in GRMHD, including physical conditions encountered in highly magnetized dynamical systems involving NSs and BHs. 
Those tests should be representative for BNS and NS-BH merger simulations, but are computationally less expensive. 
Additionally, we also carry out basic 1D shock tube tests with known solution up to high magnetization.
These results can be found in Appendix \ref{STtests}.

Our tests are employing simple analytic EOS, but we stress that the RePrimAnd scheme is completely agnostic to the 
type of EOS, provided that it respects thermodynamic and causality constraints.
This is not true for the implementation of the Noble scheme in {\tt Spritz}, which contains different 
code path depending on the type of EOS. The comparisons between the two schemes will therefore not necessarily 
generalize to any EOS. 
For all tests, we employ PPM reconstruction method, HLLE flux solver, and RK4 method for time-stepping 
with a Courant factor of 0.25.
Unless stated otherwise, we evolve the vector potential using the algebraic gauge and no Kreiss-Oliger dissipation 
\cite{Kreiss1973} (different choices are discussed in \Sref{sec:fmdisk}), and adopt BSSNOK formulation for spacetime evolution.

%=====================================================================
\subsection{Neutron star with internal magnetic field}
\label{sec:magtov}
\begin{figure}[t]
  \begin{center}
    \includegraphics[width=1\columnwidth]{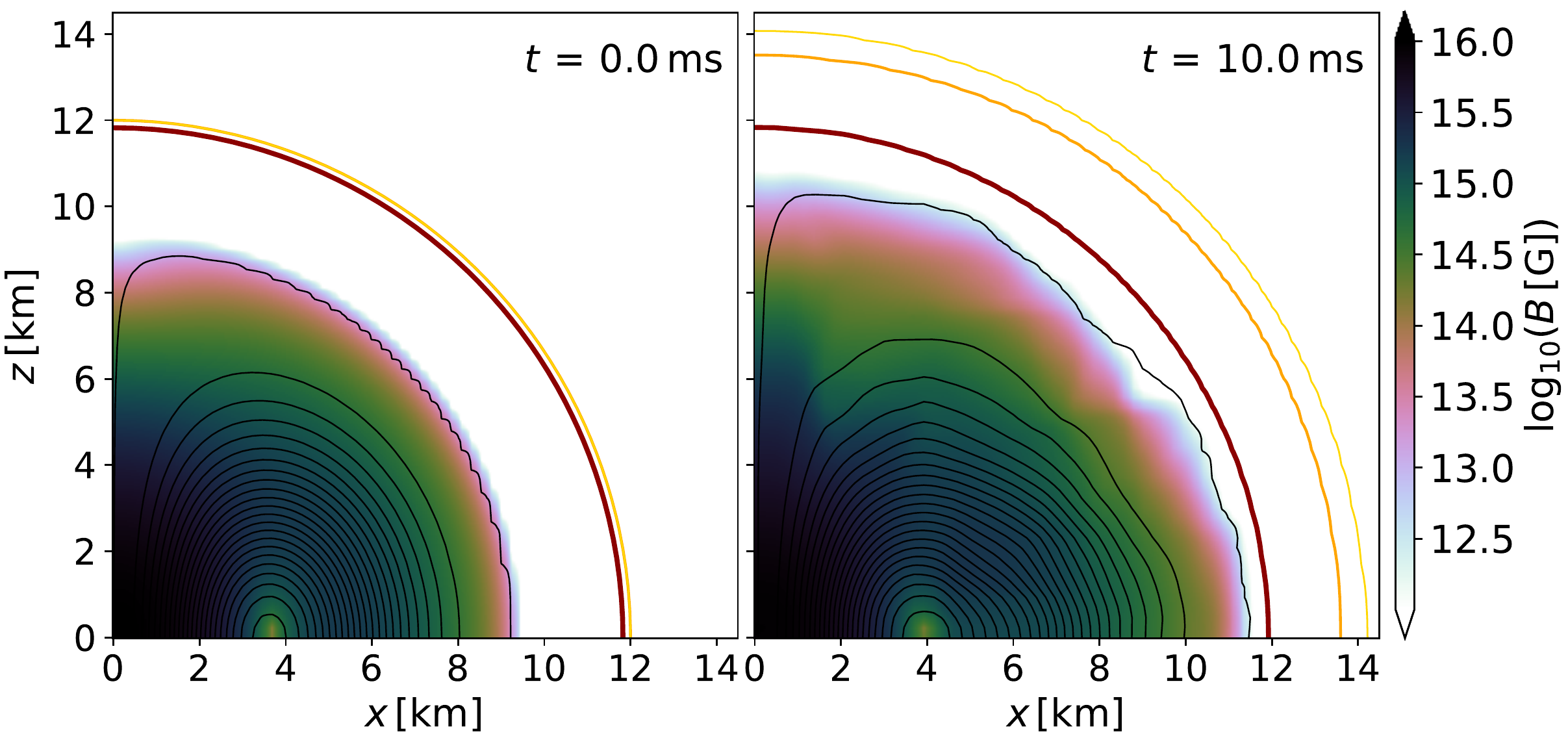}  
    \caption{Meridional snapshots of the evolution of a magnetized nonrotating NS, where the RePrimAnd C2P scheme with ideal gas EOS is adopted. Black lines correspond to $A_\phi$ isocontours and reveal the geometry of the poloidal component of the magnetic field. The red line is the iso-density contour corresponding to 1\% of the initial maximum density, whereas orange and yellow lines are the iso-density contours corresponding 100, and 5 times the artificial atmosphere density, respectively. The color scale indicates the magnetic field strength. 
    }
    \label{fig:MagTOV_Bfield2D}
  \end{center}
\end{figure}

\noindent Our first test consists of evolving a nonrotating NS endowed with initial magnetic field confined to the NS interior.
The test setup is the same as the one adopted in \cite{Cipolletta2020}. 
To obtain initial data, we first solve the Tolman-Oppenheimer-Volkoff (TOV) equations \cite{Oppenheimer:1939,Tolman:1939} which describe
the structure of a non-magnetized, static and spherically symmetric compact star in general relativity. 
As EOS, we use a polytrope $P(\rho) = K \rho^\Gamma$, with adiabatic index $\Gamma = 2$ and polytropic constant $K=100$ (in geometric units). 
We consider a model with a central rest-mass density of $\rho=8.06\times10^{14}$\,g/cm$^3$. 
This corresponds to a star with circumferential radius of about 12~km and a gravitational mass of about $1.4\,M_\odot$. 

The magnetic field is then imposed onto the TOV configuration using an analytical prescription for the vector potential.
The expression below corresponds to a field with dipolar structure confined inside the star:
\begin{align}
\label{tovavec}
\begin{split}
A_r &= A_\theta = 0 \\ 
A_\phi &=  A_\mathrm{b} \varpi^2 {\rm max} \left( p - p_\mathrm{cut}, 0 \right)^{n_s},
\end{split}
\end{align}
where $A_b$ is a scaling factor, $\varpi$ denotes the cylindrical radius, $n_s=2$ represents the degree of differentiability of the field strength, and $p_\mathrm{cut} = 0.04 p_\mathrm{max}$ establishes the cut-off pressure, confining the magnetic field entirely inside the NS and within the region where gas pressure is $\geq4\%$ of the maximum value $p_\mathrm{max}$. 
We chose $A_b$ such that the maximum magnetic field strength in the Eulerian frame is $10^{16}~\mathrm{G}$.
While this field is rather strong compared to known magnetars, it is still not strong enough to significantly deform the star, as  the magnetic energy is much smaller than the NS binding energy. We therefore leave matter distribution and spacetime metric unchanged.

We perform simulations for three different grid resolutions, i.e. low (LR), medium (MR) and high resolution (HR) using $(64)^3$, $(128)^3$ and $(256)^3$ grid-cells with grid spacing dx=$230.5$\,m, dx=$461$\,m and dx=$922$\,m, respectively, in a uniform domain with $x$-, $y$- and $z$-coordinates lying in the range $[-29.5,29.5]$\,km.
Each simulation covers 10\,ms of evolution employing the RePrimAnd primitive recovery scheme, and an atmosphere density floor set to $6.3\times10^{6}$\,g/cm$^3$. 

For the evolution, we employ two EOS implementations compatible with the initial data. 
One is the classical ideal gas defined by 
\begin{align}
P(\rho,\epsilon) &=\rho\epsilon(\Gamma - 1), 
\end{align}
using $\Gamma=2$. 
The other is the hybrid gamma-law EOS defined by 
\begin{align}
P(\rho,\epsilon) &= P_c(\rho) + (\epsilon - \epsilon_c(\rho)) \rho (\Gamma_\text{th} - 1),
\end{align}
where we use the initial data polytropic EOS for the zero-temperature curves of pressure, $P_c(\rho)$, and 
specific energy, $\epsilon_c(\rho)$, and further set $\Gamma_\text{th} = \Gamma =2$.

Evaluating the two expressions yields exactly the same EOS function $P(\rho,\epsilon)$, but 
there is a subtle difference regarding the definition of temperature and the validity range.
When specifying an EOS as $P(\rho,\epsilon)$, there is an ambiguity about the definitions of entropy and temperature. 
We recall that for the classical ideal gas, the pressure and specific internal energy are proportional 
to the temperature, and the lower validity bound given by zero temperature is $\epsilon \ge 0$. 
For the hybrid EOS, the zero temperature curve is the polytrope and the lower validity bound
is $\epsilon \ge \epsilon_c(\rho)$.
The RePrimAnd C2P scheme enforces those validity ranges for ideal gas and hybrid EOS, which 
might introduce small differences between simulations using the two. The Noble scheme on the other hand
does not enforce EOS validity ranges.

To assess the consequences, we
perform simulations of the same initial data using either the ideal gas or the equivalent hybrid EOS. 
As long as the C2P functions correctly, we expect identical results for the Noble scheme and RePrimAnd in
conjunction with the classical ideal gas EOS (assuming that the specific energy might drop below the 
polytrope but still remains positive). This point is relevant because not 
all types of EOS can be evaluated outside the physically meaningful range, most notably fully tabulated EOS. 
Without an EOS-agnostic policy regarding validity ranges, C2P schemes might behave differently for different 
types of EOS. 

\begin{figure}
  \begin{center}
    \includegraphics[width=1.0\columnwidth]{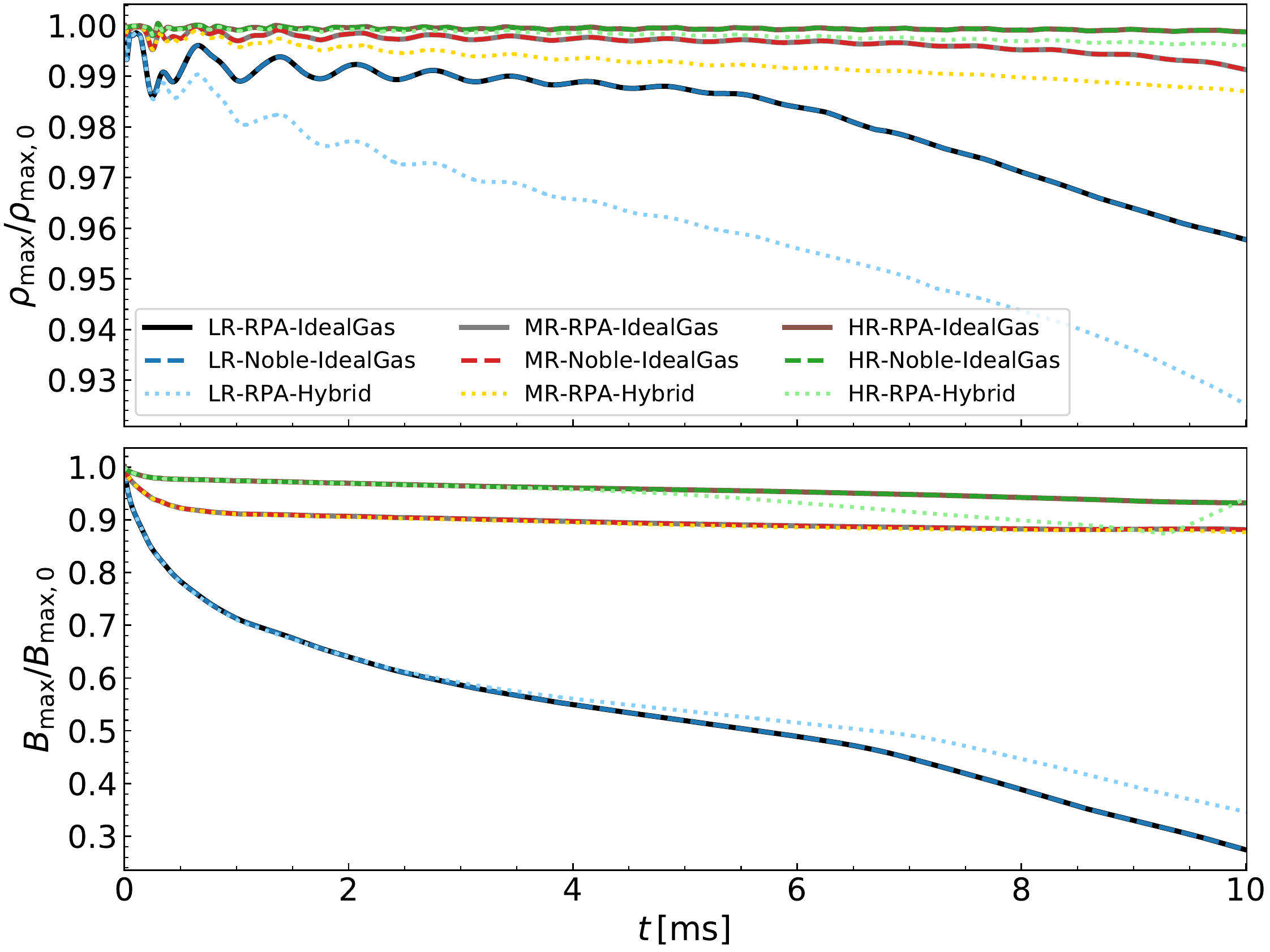}  
    \caption{Evolution of normalized maximum rest-mass density (top) and magnetic field strength (bottom) for 
    low (LR), medium (MR), and high resolution (HR) simulations, comparing also between the Noble C2P (dashed 
    lines) scheme and the RePrimAnd (RPA) scheme, the latter either in conjunction with the ideal gas EOS (solid 
    lines) or the hybrid EOS (dotted lines).}
    \label{fig:MagTOV_Rho_Bmax}
  \end{center}
\end{figure}

\Fref{fig:MagTOV_Bfield2D} shows snapshots of initial and final magnetic field strength and density contours, for the HR run with the RePrimAnd scheme. 
The low-density region at the surface is undergoing an expansion due to spurious heating, which is a typical 
behavior for evolution schemes near NS surfaces. In 10\,ms, the magnetic field configuration also shows some variations, in particular in the outer portion of the NS (radii above $\simeq\!6$\,km). We note that the simulation time span is of the order of the Alfv\'en timescale at this field strengths and thus field readjustments can occur, possibly triggered by the know instability of purely poloidal fields (e.g., \cite{Ciolfi2011} and refs.~therein). 
The evolution of the normalized maximum rest-mass density $\rho_\mathrm{max}$ and maximum magnetic field strength $B_\mathrm{max}$ is reported in the top and bottom panels of \Fref{fig:MagTOV_Rho_Bmax}, respectively, for all three resolutions.
\begin{figure}
  \begin{center}
    \includegraphics[width=1\columnwidth]{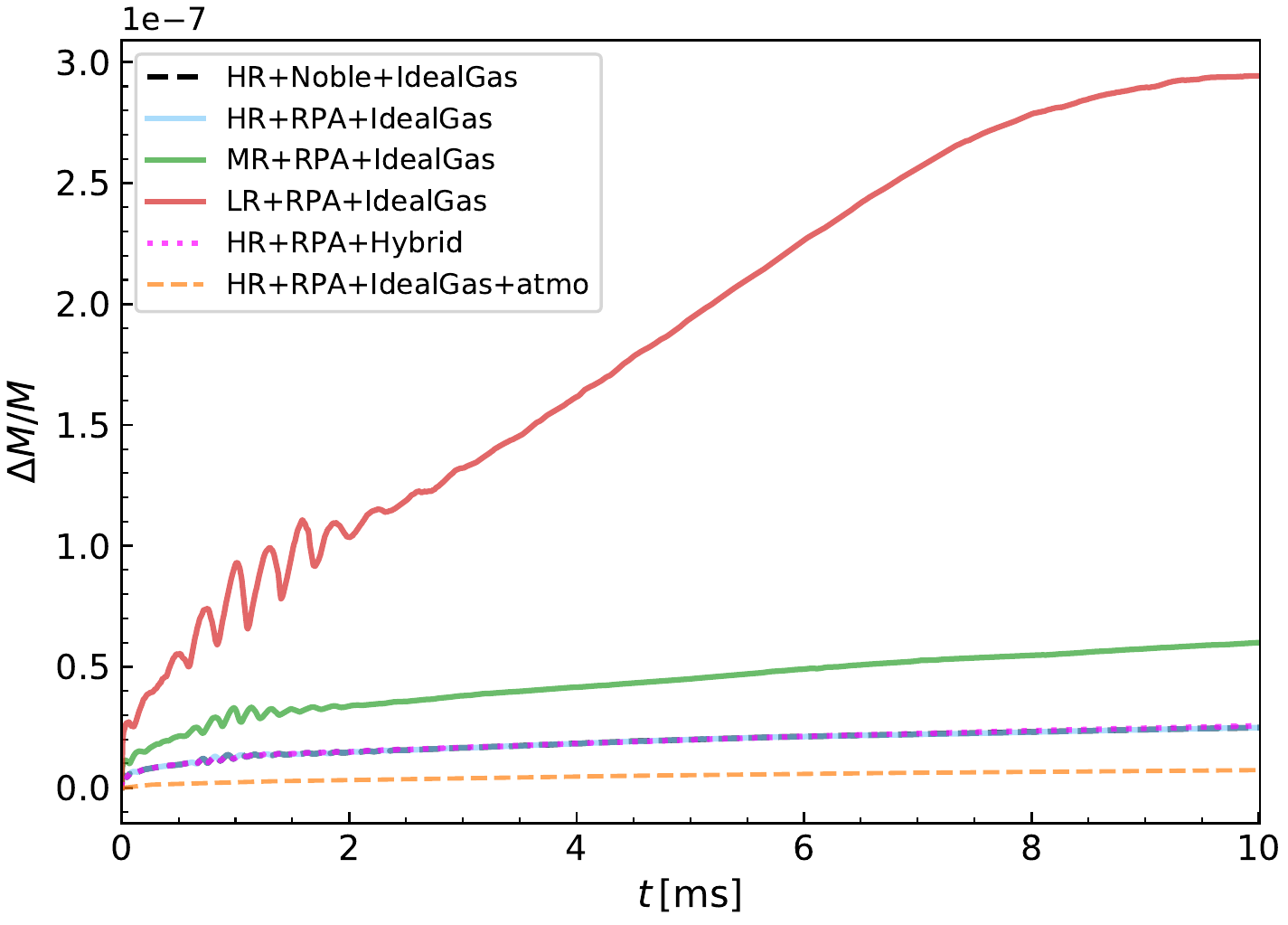}  
    \caption{Relative change in total baryonic mass for the different tests with Noble and RePrimAnd schemes. 
    }
    \label{fig:MagTOV_mass_cons}
  \end{center}
\end{figure}
The maximum density $\rho_\mathrm{max}$ shows oscillations which are decaying while the system settles toward
a numerical equilibrium. As we will show later, those oscillations correspond to oscillation modes expected from
perturbation theory. With increasing resolution, the initial excitation amplitude decreases.  In addition, $\rho_\mathrm{max}$ and $B_\mathrm{max}$ exhibit a continuous drift, which we attribute to numerical evolution errors (compare also \cite{Baiotti2005}). 
With increasing resolution, such a drift clearly decreases. 

To assess the overall convergence behavior, we consider the time evolution of maximum density, maximum magnetic field strength, and minimum lapse function. Using the $L_2$ norms over time of the differences between high and medium resolution and between medium and low resolution, 
we find an average convergence order around 2.5 for the density, 2.7 for the magnetic field, and 3.1 for the lapse. 
At the finite resolutions used here, this is typical behavior for such simulations, which employ numerical schemes 
with different convergence orders for spacetime and magnetohydrodynamic parts.

Comparing the results obtained with the Noble scheme and the RePrimAnd scheme in conjunction with the ideal gas evolution EOS, we observe an exact match. Both schemes are equally capable to handle this test case. 

Comparing between the RePrimAnd scheme with ideal gas and equivalent hybrid EOS, we find a somewhat larger drift for the latter.
The only difference in the evolution is that for the hybrid EOS, the specific energy is mapped back to the value given by the polytropic cold EOS when evolution errors cause it to fall below. Hence, the correction only increases the internal energy. We conclude that on average,
enforcing the validity range slightly increases the spurious heating. Enforcing the validity range might seem like a drawback, but should be weighted
against consistent behavior for any type of EOS.  

Since the artificial atmosphere is also handled by the C2P implementations, it is important to validate the 
conservation of total baryonic mass. \Fref{fig:MagTOV_mass_cons} shows the total baryonic mass, comparing the
different resolutions for the RePrimAnd case, the RePrimAnd and Noble schemes at high resolution, and the 
RePrimAnd case with ideal gas and hybrid EOS. For completeness, we compare two different measures for the mass, 
including and excluding the artificial atmosphere contribution. Both the Noble and RePrimAnd implementations
meet the expectations for mass conservation, and agree exactly for the case of ideal gas (for all 
resolutions). The enforcing of validity bounds for the hybrid EOS does not have an impact on mass conservation, as expected.

\begin{figure}
  \begin{center}
    \includegraphics[width=1\columnwidth]{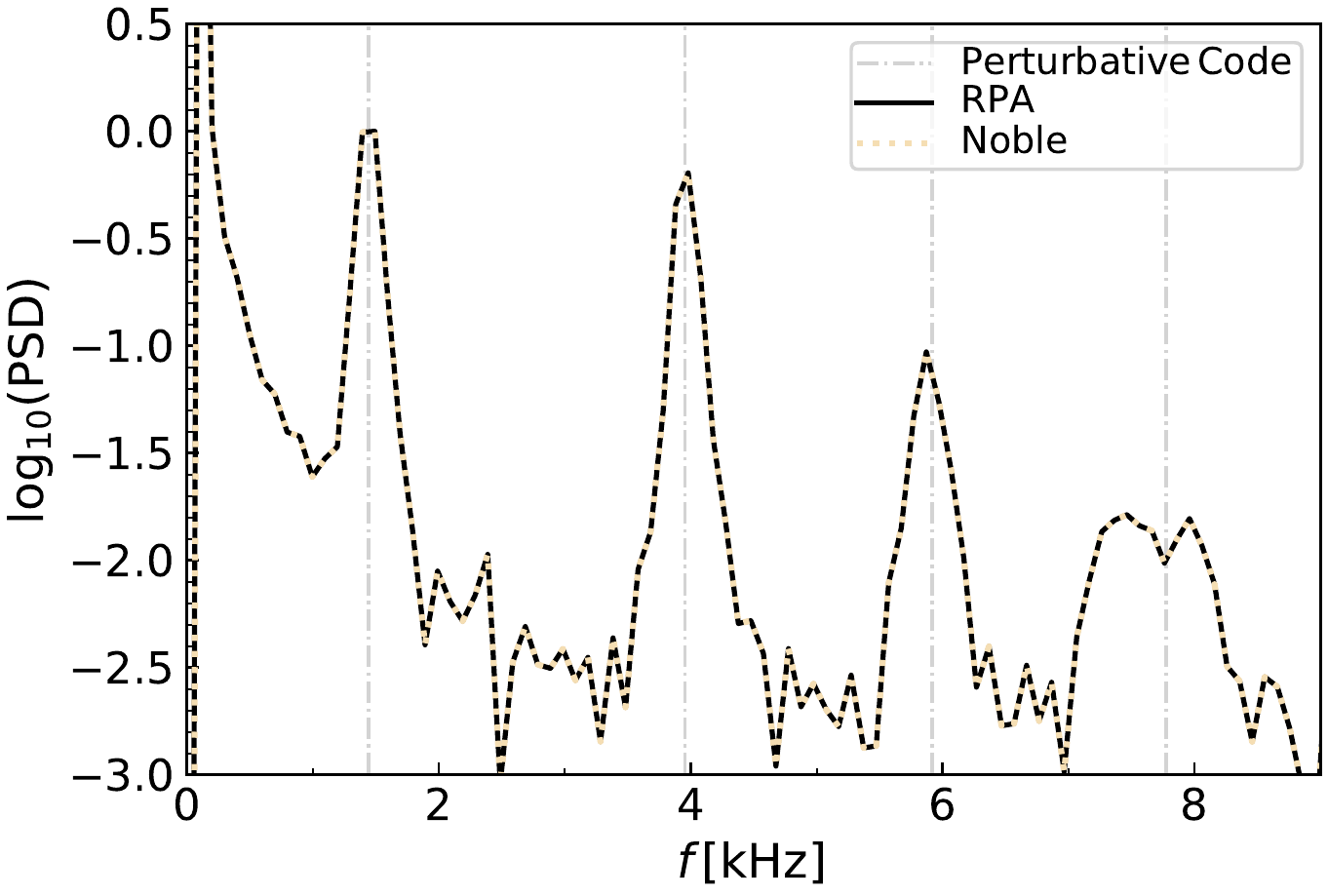}  
    \caption{Fourier transform of the maximum rest-mass density evolution normalised to the amplitude of first frequency peak, for both the RePrimAnd and Noble schemes. The dash-dotted vertical lines represent the normal mode frequencies computed independently via a perturbative code as reported in \cite{Font2000}.
    }
    \label{fig:MagTOV_OscFreq}
  \end{center}
\end{figure}

In \Fref{fig:MagTOV_OscFreq}, we plot the power spectrum of the maximum rest-mass density evolution.
As one can see, the spectra obtained with the RePrimAnd and Noble schemes are identical.
Further, the spectrum shows clear peaks that are located at frequencies of the radial NS oscillation modes,
predicted in \cite{Font2000} using a 2D perturbative code. The oscillations of maximum density visible in 
\Fref{fig:MagTOV_Rho_Bmax} thus correspond to NS oscillations excited by the deviation of initial data from equilibrium.

Finally, we assess the computational costs of the two C2P implementations. Using the timing functionality 
available in the {\tt Einstein Toolkit}, we find that both schemes use about $2.5\%$ of the total computational time.
For the test at hand, the C2P is not a computational bottleneck of the \texttt{Spritz} code, and the computational efficiency of both schemes is similar.
We note that the fraction spent in the C2P might increase significantly when using a much slower type of EOS, for 
example an interpolation table using temperature as one independent variable (see also the discussion in \cite{Kastaun2021}).
\begin{figure*}[t]
  \begin{center}
    \includegraphics[width=2.05\columnwidth]{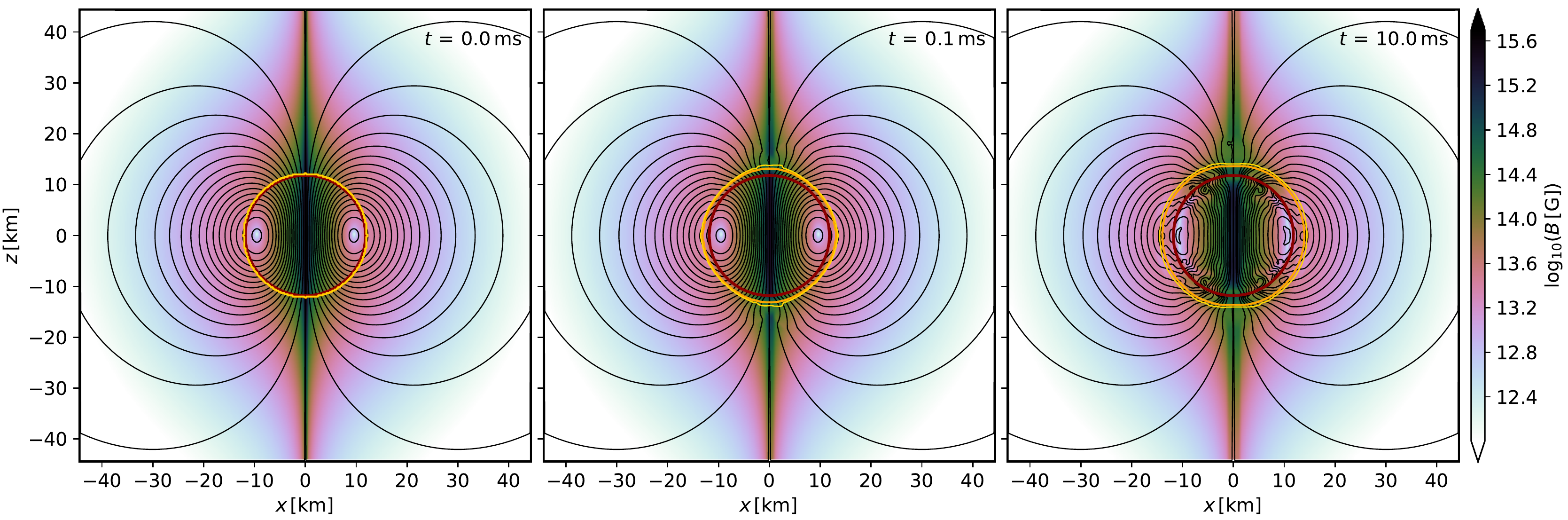}  
    \caption{Results for the test case of a nonrotating NS endowed with an external dipolar magnetic field, evolved with the RePrimAnd C2P scheme and using a low artificial atmosphere density of $6.3\times10^{5}$\,g/cm$^3$. The black lines are isocontours of the vector potential component $A_\phi$. Rest mass density contours in red, orange and yellow are illustrated in the same fashion as done in Fig.~\ref{fig:MagTOV_Bfield2D}.
    }
    \label{fig:DipoleTOV_Kastaun_Bfield2D}
  \end{center}
\end{figure*}
\begin{figure*}
  \begin{center}
    \includegraphics[width=2.0\columnwidth]{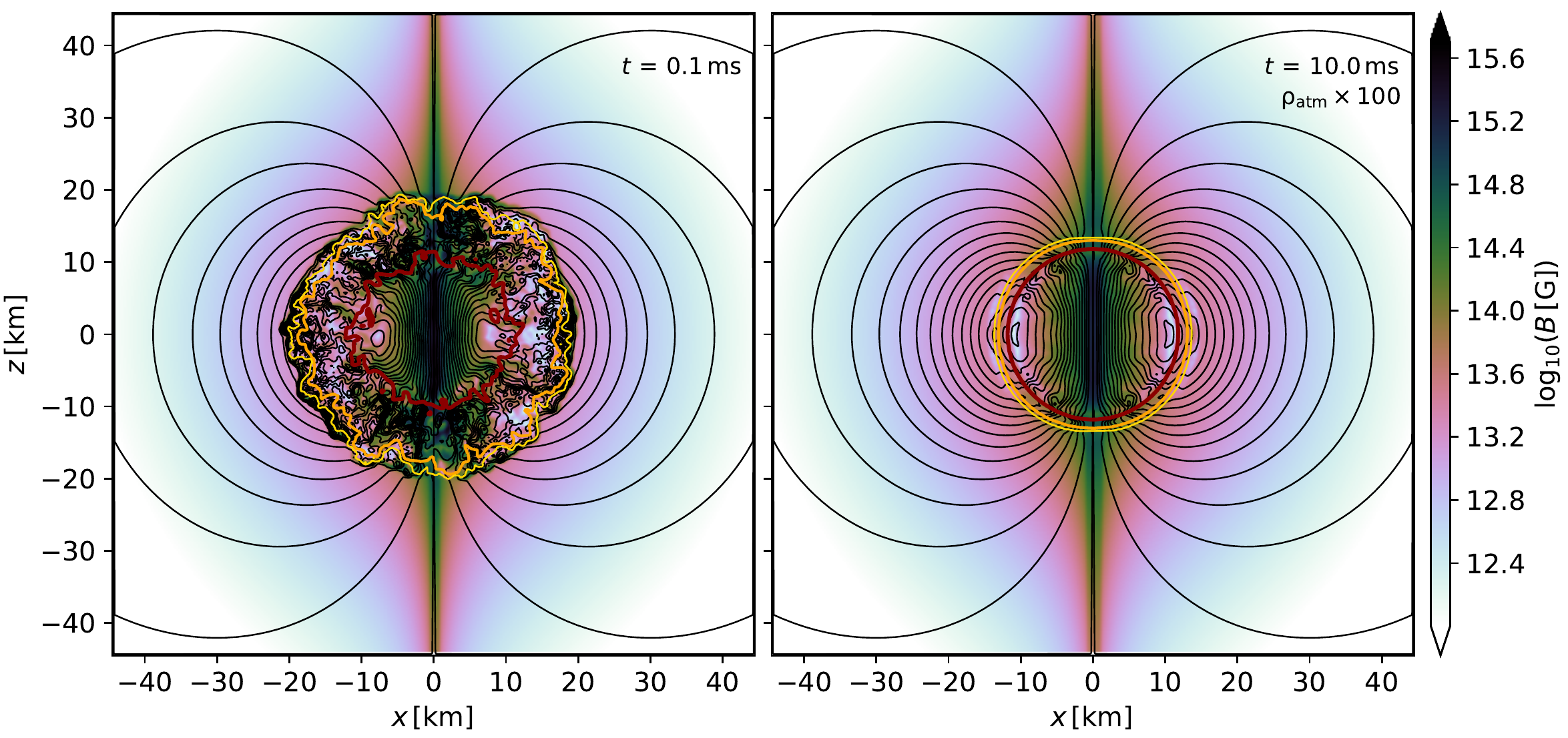}  
    \caption{Results for the external dipole test obtained with the Noble C2P scheme. 
    Left: Same as the middle panel of Fig.~\ref{fig:DipoleTOV_Kastaun_Bfield2D}, 
    but showing the instability developing when adopting the Noble scheme. 
    Right: Snapshot after $10$ ms evolution with the Noble C2P scheme when using an artificial 
    atmosphere density of $6.3\times10^{7}$\,g/cm$^3$, enlarged by a factor 100 compared 
    to the RePrimAnd simulation results shown in the right panel of Fig.~\ref{fig:DipoleTOV_Kastaun_Bfield2D}.
    }
    \label{fig:DipoleTOV_Noble_Bfield2D_diffAtmo}
  \end{center}
\end{figure*}

%=====================================================================
\subsection{Neutron star with extended dipolar magnetic field}
\label{sec:dipoletov}

\noindent A critical aspect for any primitive variable recovery scheme for MHD is the ability to handle high magnetizations in low density regimes. 
While the RePrimAnd scheme has proven to be robust and reliable when dealing with extremely high magnetisations in pointwise (i.e.~single fluid element) tests (see \cite{Kastaun2021} for details), more delicate dynamical tests involving high magnetizations together with a nontrivial global evolution represent the next necessary step for a full validation in view of future BNS merger simulations. 

Here, we simulate the same nonrotating NS model as reported in \Sref{sec:magtov}, but now endowed with a dipole magnetic field also extending outside its surface. In detail, the initial magnetic field is set up employing a vector potential of the form (also used in \cite{Moesta2020})
\begin{align}
\label{tovavecExt}
\begin{split}
A_r &= A_\theta = 0,\\ 
A_\phi &=  B_0 \frac{r^3_0}{r^3 + r^3_0} r \sin \theta ,
\end{split}
\end{align}
where $B_0\!=\!10^{15}$\,G (corresponding to a maximum magnetic field strength of $\simeq6.6\times10^{15}$\,G), and we assume $r_0=12$\,km. 

We perform a simulation using the RePrimAnd scheme together with the ideal gas EOS. 
The grid domain extends up to 44.3\,km in all cartesian directions, with grid spacing of $dx\simeq231$\,m (corresponding to the high resolution introduced in the previous Section). In order to make the test more demanding, we chose a very low artificial atmosphere density of $6.3\times10^{5}$\,g/cm$^3$. 

We evolved the system up to $10$\,ms without encountering any problems. 
Most importantly, the RePrimAnd C2P scheme worked robustly even in the low-density regions outside the NS, where the magnetic-to-fluid pressure ratio $\beta^{-1}$ reaches large values up to $\sim10^4$.
\Fref{fig:DipoleTOV_Kastaun_Bfield2D} shows three snapshots of the magnetic field strength in the meridional plane, along with vector potential and rest-mass density contours.
Apart from some small drift due to normal evolution errors, the field structure remains stable.

When evolving the same configuration using the Noble scheme, the simulation develops strong instabilities after only $\simeq0.1$\,ms. These instabilities first appear shortly outside the NS surface in the polar region, where the  
magnetic field strength is largest. Subsequently, they spread around the rest of the star.
As shown in the left panel of \Fref{fig:DipoleTOV_Noble_Bfield2D_diffAtmo}, the magnetic field in the outer layers of the NS becomes highly disordered and even the density isocontours reveal a very complex, unexpected dynamics that ultimately leads to failures in spacetime evolution. 
The exact mechanism behind the instability is unknown. One possible explanation is that the inaccuracies 
of the Noble C2P solution might be larger in the parameter space of the affected regime and then interact 
with the evolution scheme in an unstable manner. Since the Noble scheme depends on initial guesses from the evolution, 
such effects would be difficult to reproduce in standalone tests such as \cite{Siegel2018c2p}.

When the atmosphere floor density is increased by a factor of 100, the simulation performs well also with the Noble scheme, as shown in the right panel of \Fref{fig:DipoleTOV_Noble_Bfield2D_diffAtmo}. 
Further tests show that increasing the atmosphere density by an intermediate factor 10 is not sufficient and still leads to an early failure. 
We therefore attribute the failure to the magnetization, which is maximal at lowest density and hence reduced by increasing the atmosphere density. 

\begin{figure}[t]
  \begin{center}
    \includegraphics[width=0.9\columnwidth]{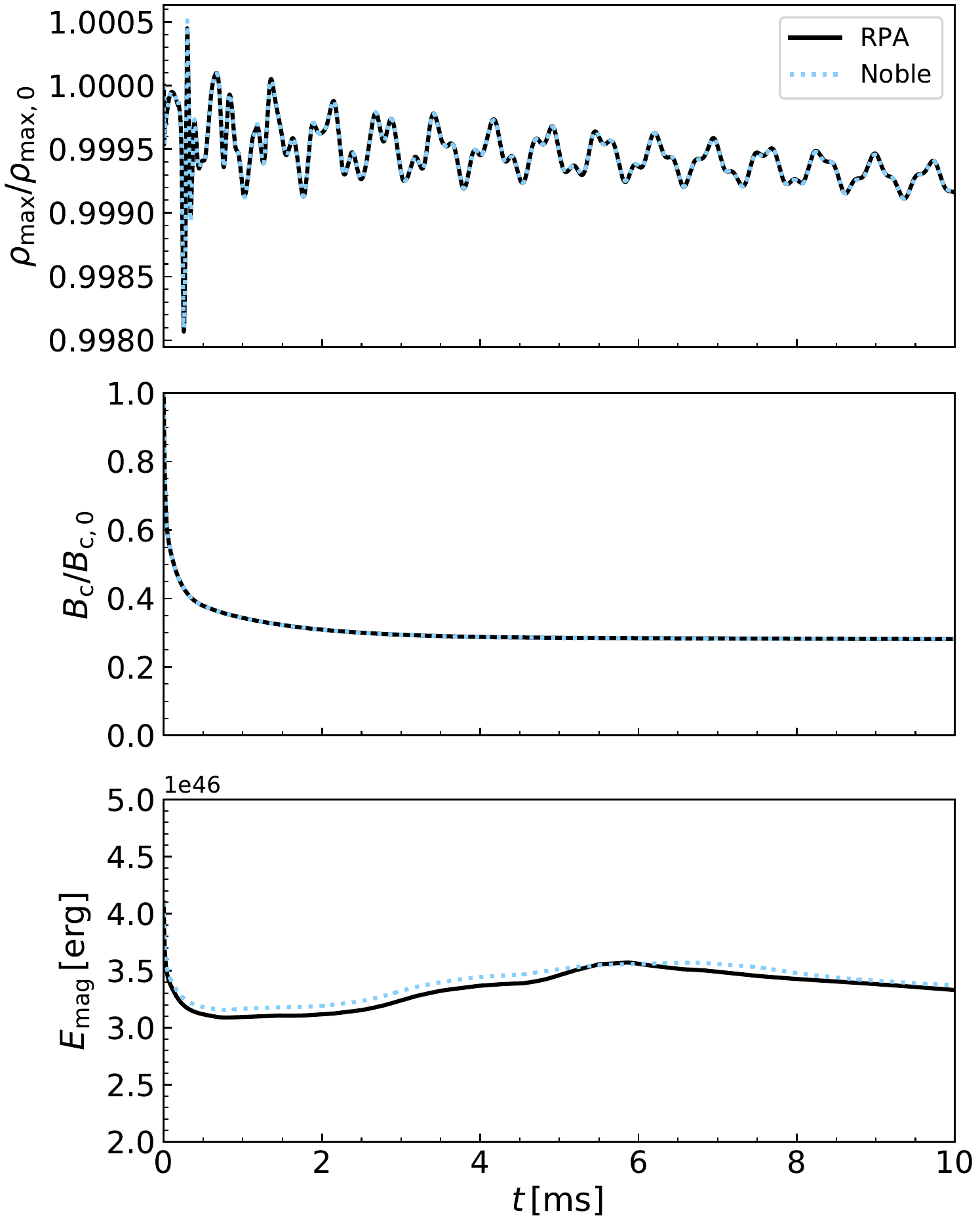}  
    \caption{Evolution of normalized maximum rest-mass density (top), normalized magnetic field strength at the center (middle), and magnetic energy (bottom) for the case of a nonrotating NS endowed with extended dipolar magnetic field, evolved with the RePrimAnd C2P scheme and artificial atmosphere density of $\rho_\mathrm{atm}=6.3\times10^{5}$\,g/cm$^3$ or with the Noble C2P scheme and $\rho_\mathrm{atm}=6.3\times10^{7}$\,g/cm$^3$.
    }
    \label{fig:DipoleTOV_Rho_Emag}
  \end{center}
\end{figure}

When using the Noble scheme with 100 times larger atmosphere, the overall results (NS surface aside) are very similar to the RePrimAnd case with the original low density atmosphere.
\Fref{fig:DipoleTOV_Rho_Emag} compares the evolution of maximum rest-mass density $\rho_\mathrm{max}$ (top panel), central magnetic field strength $B_\mathrm{c}$ (middle panel), and total magnetic energy (bottom panel) for the two cases.

The tests clearly show that the RePrimAnd C2P scheme increases the robustness of the \texttt{Spritz} code when dealing with magnetic-to-fluid pressure ratios at least up to $\beta^{-1} \sim 10^4$, compared to the Noble scheme used previously. We stress that our study is not testing the accuracy 
of the numerical evolution in highly magnetized low-density regions, and point out that evolving such regions more accurately 
might require specialized evolution schemes incorporating  the force-free approximation and/or resistive MHD. However, 
it is already a big advantage that the presence of such regimes is not causing a failure of the whole simulation.

%=====================================================================
\subsection{Rotating magnetized neutron star}
\label{sec:rotns}

\noindent Moving to a more demanding test, we now consider a uniformly rotating NS with magnetic field extending outside the star. The initial data are generated using the {\tt Hydro\_RNSID} module of the Einstein toolkit. Our model employs a $\Gamma=2$ polytropic EOS with $K=100$ (in units $c=G=M_\odot=1$), a central density $\rho_{\rm c}\simeq6.87\times10^{14}$\,g/cm$^3$ and a polar-to-equatorial radius ratio of 0.8744. 
This produces a uniformly rotating NS on the stable branch with a gravitational mass of 1.4\,$M_\odot$, an equatorial radius of 15.65\,km, and a rotation rate of 500\,Hz.

We add an extended dipolar magnetic field on the NS in the same way as done in \Sref{sec:dipoletov}. 
The magnetic dipole axis coincides with the rotational axis.
Different to \Sref{sec:dipoletov}, we do not add the magnetic field to the initial data, but evolve the non-magnetized initial data for 10\,ms before imposing the magnetic field. 
The motivation for this choice is to obtain a test case that is representative for the application to binary neutron star mergers 
using an approach similar to, e.g., \cite{Ruiz2016}, where the extended magnetic field is added to the NSs during the last 
orbits before merger.

After 10\,ms of evolution without magnetic field, the NS is surrounded by a hot, low-density cloud of matter.
Such diffuse halos are a typical effect of perturbations near the NS surface initially triggered by numerical evolution errors. This expansion of the outer layers, along with the interaction with the artificial atmosphere, also leads to strong differential rotation in the corresponding regions.
For the actual test, we then evolve the magnetized system for another 10\ ms.
The atmosphere floor density is again set to a low value of $6.3\times10^{5}$\,g/cm$^3$.

For this test, we also use `box-in-box' mesh refinement with four refinement levels.  The coarsest(finest) computational grid extends up to $\simeq591(74)$\,km in all Cartesian directions, and the finest grid spacing is $dx=461$\,m, which is twice as large as the grid spacing used in the previous test. This setting gives about 26 grid points along the NS radius.
The test allows us to validate the correct interplay between the implementation of the new C2P scheme in \texttt{Spritz} and the mesh refinement. 

\begin{figure*}
  \begin{center}
    \includegraphics[width=1.6\columnwidth]{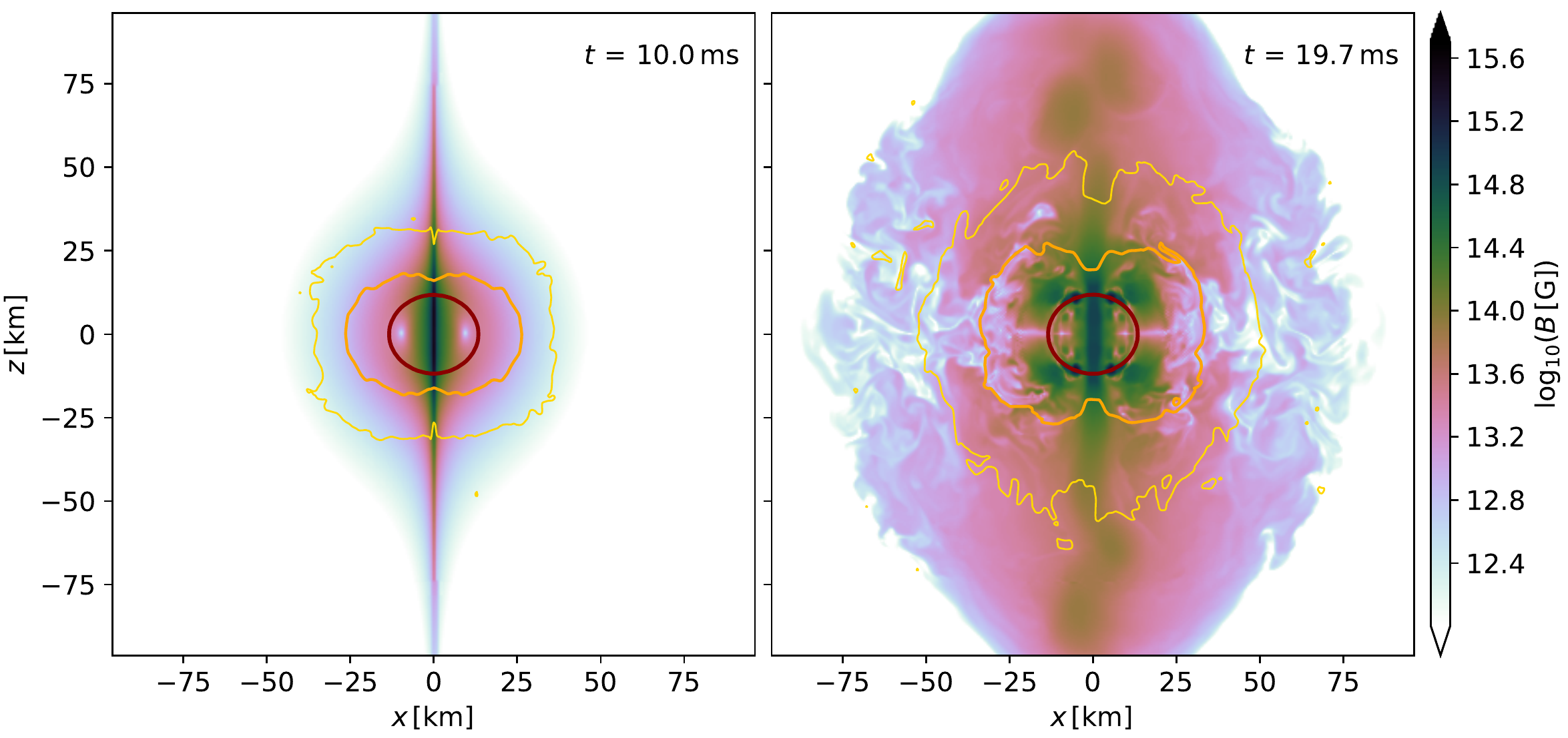}  
    \caption{Meridional view of a magnetized rotating NS, evolved using the RePrimAnd C2P scheme. Color scale indicates the magnetic field strength and the iso-density contours are the same as in Fig.~\ref{fig:MagTOV_Bfield2D}. Left panel shows the time at which an extended dipolar magnetic field is imposed (10\,ms), while the right panel shows the system towards the end of the simulation.
    }
    \label{fig:rotns_bfield2d}
  \end{center}
\end{figure*}

\Fref{fig:rotns_bfield2d} shows the magnetic field strength on the meridional plane along with selected isocontours of rest-mass density, both at 10\,ms (time at which the field is added) and 20\,ms (final simulation time).
Right after the field is imposed, it starts to get distorted by the motion of the low density material surrounding the NS bulk. Differential rotation leads to magnetic winding and the appearance of a toroidal component. The resulting magnetic pressure gradients contribute to the expansion of the outer layers of the NS and of the surrounding matter. 

\begin{figure}
  \begin{center}
    \includegraphics[width=1\columnwidth]{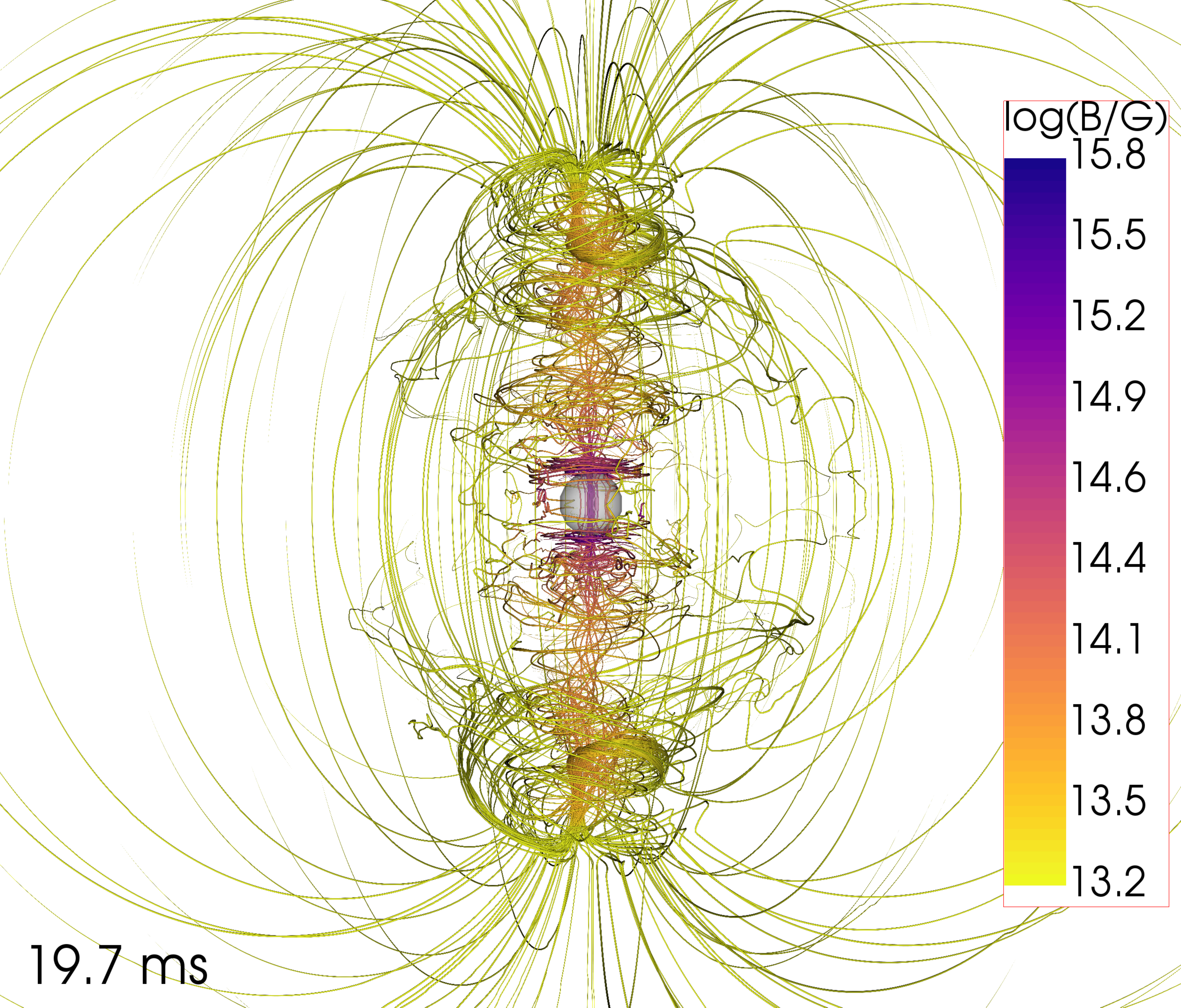}  
    \caption{Three dimensional magnetic field structure of the system shown in right panel of Fig.~\ref{fig:rotns_bfield2d}. A sphere of radius 10\,km is added in the center to give a spatial scale reference.
    }
    \label{fig:rotns_bfield3d}
  \end{center}
\end{figure}

In \Fref{fig:rotns_bfield3d}, we plot the field line structure close to the star towards the end of the simulation, 
using the visualization method developed in \cite{Kawamura2016} but focusing more on the regions near the NS and 
rotation axis.
Along the spin axis and up to about 100\,km from the center, we observe the twisting of magnetic field lines\footnote{Such an helical structure is also slightly bent, hinting towards the development of a kink instability (e.g., \cite{Kiuchi2012}). In-depth analysis of this effect is however beyond the scope of this work.}. 
At large distances, within the artificial atmosphere, the original dipolar structure remains frozen (see Sec.~\ref{sec:formulation}). We stress that a proper treatment of those outer regions would require a force-free evolution, which is not currently implemented in {\tt Spritz}. 

During the whole simulation, the RePrimAnd C2P scheme experienced no difficulties while handling the high magnetizations present in the low density regions, where the magnetic-to-fluid pressure ratio reaches $\sim\!10^4$. 

%=====================================================================
\subsection{Rotating magnetized neutron star collapse}
\label{sec:collapse}

\noindent Our next test case consists of evolving the collapse of a magnetized differentially rotating hypermassive NS to a BH.
Even though the gauge conditions avoid the physical singularity, the center of BHs is challenging for numerical evolution schemes. 
The purpose of the test is not just to demonstrate the robustness of the C2P scheme under the conditions encountered during BH formation, but also to validate that the interplay between our policy for corrections and the time evolution scheme results in a stable evolution of a BH with matter.
As described in Sec.~\ref{sec:formulation}, we evolve the BH interior without excision, but employ a C2P policy near the center that effectively limits the velocity and specific energy.
\begin{figure*}
  \begin{center}
    \includegraphics[width=2\columnwidth]{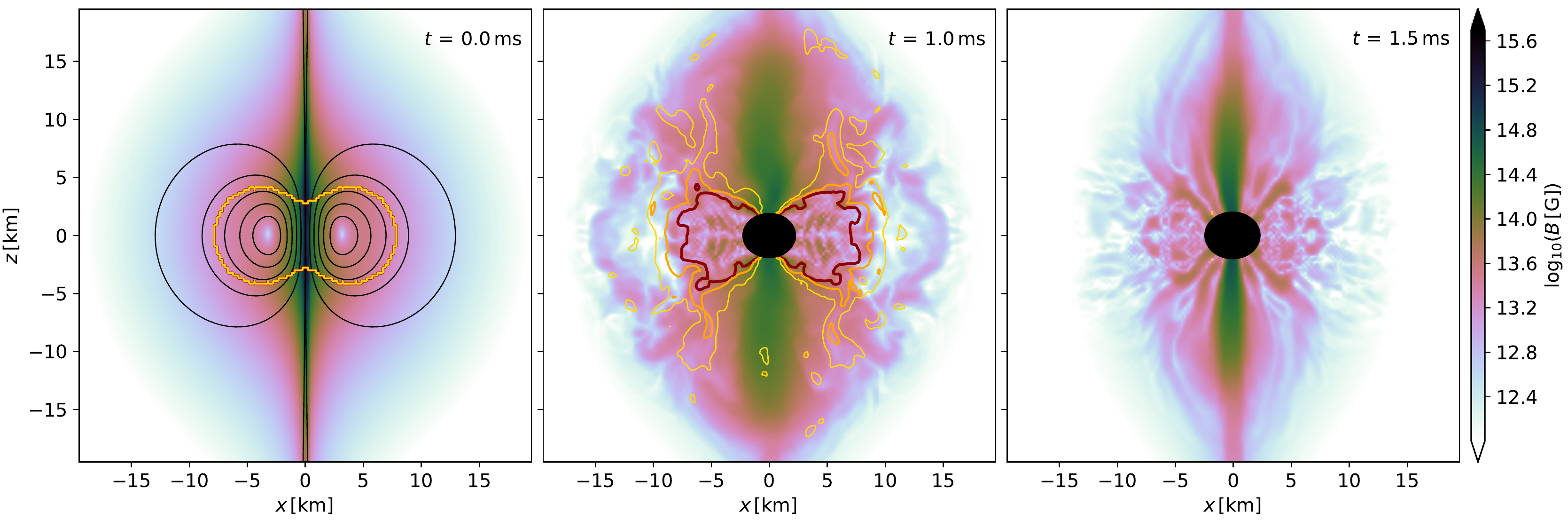}  
    \caption{Meridional snapshots of magnetic field strength for the magnetized rotating NS collapse simulation, evolved using the RePrimAnd C2P scheme. The iso-density contours in red, orange, and yellow correspond to 5, 3 and 1.5 times the artificial atmosphere density, respectively. 
    The left panel, referring to the initial setup, shows also the magnetic field geometry as $A_\phi$ isocontours (black). The middle and right panels refer to 1 and 1.5\,ms of evolution, respectively. The $A_\phi$ isocontours are not shown in the last two panels as their disordered structure overcrowds the regions of interest we want to show. The black area marks the apparent horizon of the BH.    
    }
    \label{fig:rotnscoll_bfield2d}
  \end{center}
\end{figure*}

For this test, we adopt the dynamically unstable axisymmetric model A2 of \cite{Giacomazzo2011}, which consists of a hypermassive NS close to collapse. It employs a polytropic EOS with $\Gamma=2$ and $K=100$. The model has a gravitational mass of $M\simeq 2.228\,M_\odot$ and a central rest-mass density of $\rho_{\rm c} \simeq 1.892\times10^{15}$\,g/cm$^3$. 
The model follows the ``j-constant law" of differential rotation,
with differential rotation parameter $\hat{A}=1$ (see \cite{Giacomazzo2011} for details)
and central angular velocity $\Omega_{\rm c}\simeq 44 \, \mathrm{Rad}/\mathrm{ms}$.
The resulting ratio of the polar to the equatorial coordinate radii is $r_{\rm p}/r_{\rm e} = 0.33$. 

The non-magnetized model is constructed using the {\tt Hydro\_RNSID} module of the {\tt Einstein Toolkit}.
On top of this initial configuration, we add an extended dipolar magnetic field in the same fashion as done in \Sref{sec:dipoletov}, with $B_0=10^{15}$\,G. However, in this case we choose $r_0=4$\,km. 

The computational grid setup is the one described in \Sref{sec:rotns}, supplemented with an additional inner refinement level 
that has a grid spacing of $dx\simeq230.5$\,m and which extends up to $[-37,+37]$\,km in all three directions. The system is evolved with \texttt{Spritz}, employing the RePrimAnd C2P scheme together with an ideal gas EOS with $\Gamma=2$. The atmosphere floor density is again set to $6.3\times10^{5}$\,g/cm$^3$.
\begin{figure*}
  \begin{center}
    \includegraphics[width=1.6\columnwidth]{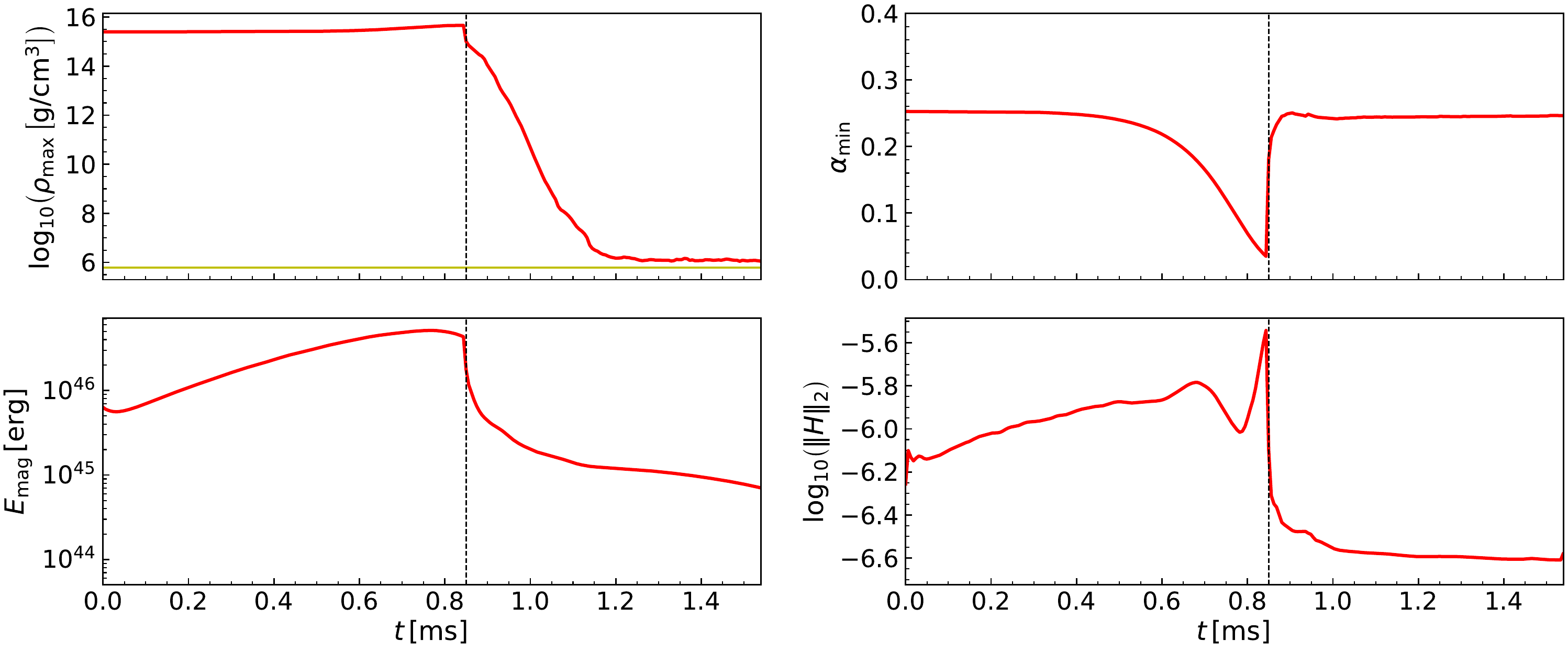}  
    \caption{Evolution of maximum rest-mass density normalized to the initial value (top-left), total 
    magnetic energy (bottom-left), minimum of lapse (top-right), and L2 norm of the Hamiltonian constraint (bottom-right). 
    Once the BH is formed, the interior of the apparent horizon is excluded for the computation of these quantities. The vertical black (dashed) line in all panels marks the time at which the apparent horizon is found in the simulation, while the horizontal yellow line in the top-left panel represents the floor density.
    }
    \label{fig:rotnscoll_1d}
  \end{center}
\end{figure*}

We evolve the system for around 1.5~ms, which is sufficient to capture the collapse of the hypermassive star and the formation of the BH. Prior to collapse (occurring at $\simeq\!0.84$\,ms), the outer layers of the NS tend to expand due to spurious heating, while the central core is slightly contracting. The initially ordered and purely poloidal magnetic field is wound up by differential rotation, which creates a toroidal component and amplifies the overall field strength. At the same time, the field becomes more disordered within the expanding outer layers of the NS, due to the irregular fluid motion. After BH formation, matter is rapidly absorbed and the magnetic field is also dragged along by the in-falling material. \Fref{fig:rotnscoll_bfield2d} illustrates the magnetic field strength at the initial, intermediate, and final times of the simulation. Throughout the simulation, we do not find any significant artifacts in magnetic field at the refinement level jumps.

The above dynamics is reflected in the time evolution of maximum rest-mass density $\rho_{\rm max}$, total magnetic energy $E_{\rm mag}$, minimum lapse $\alpha_{\rm min}$, and $L_2$-norm of the Hamiltonian constraint $H$, which are shown in \Fref{fig:rotnscoll_1d}.
Note that once the BH has formed, the interior of the apparent horizon is excluded from those measures.

We conclude that the RePrimAnd scheme can be used for scenarios involving BH formation, most notably BNS and NS-BH mergers.
Moreover, the robustness and flexibility in applying corrections allows to evolve BHs in GRMHD schemes without applying
any form of excision in the center of the BH.

%=====================================================================
\subsection{Fishbone-Moncrief BH-accretion disk}
\label{sec:fmdisk}

For our final 3D test of the RePrimAnd scheme, we simulate a magnetized accretion disk surrounding a Kerr BH, based on the equilibrium torus solutions proposed by \cite{fishbonemoncrief1976}. 
A BH-disk system is one of the typical outcomes of BNS/NS-BH mergers (e.g., \cite{Paschalidis2015, Ciolfi2017,Foucart2019, Shibata2019, Bernuzzi2020, Bernuzzi2020remnants, Ruiz2020, Kastaun2021numerical}) and model investigations of such systems probe the underlying mechanisms that govern the dynamics of the plasma accretion, post-merger disk winds that contribute to kilonova emission, as well as launching of relativistic jets. Such a system has also been widely used as a standard test case for different evolution codes and schemes \cite{HARM2003,Noble2006,HARM3D2009,Porth2017,Janiuk2018,Ripperda2019,Berthier2021}. In our case, evolving such a configuration allows another crucial check on the robustness of the RePrimAnd scheme when dealing with accretion onto a BH and MHD turbulence. 

The initial data for spacetime and hydrodynamic variables is built using the {\tt FishboneMoncriefID} thorn of the {\tt Einstein Toolkit} \cite{EinsteinToolkit1, EinsteinToolkit2}, which has been adapted to work with {\tt Spritz} . The public version of thorn was also used, e.g., to setup the {\tt IllinoisGRMHD} \cite{etienne2015illinoisgrmhd} simulations in the code comparison reported in \cite{porth2019event}.
For the initial data, we use a polytropic EOS with $\Gamma=4/3$ and $K \simeq 0.156$, while for the evolution we use an ideal gas EOS with $\Gamma=4/3$.
We set the initial Kerr BH mass to $2.7 \, M_\odot$ and its dimensionless spin parameter to $a=0.9$. This results in an equatorial BH horizon radius of about 7.8\,km in simulation coordinates. The inner edge of the torus is located at a coordinate radius $r_{in} \approx 24 \,$km, and the maximum density (9.76$\times10^{11}$\,g/cm$^3$) is reached at coordinate radius $r_{max} \approx 48 \,$km. This matter configuration is then complemented by a purely poloidal magnetic field confined inside the torus, using expression \Eref{tovavec}, and choosing an initial maximum magnetic field strength of $10^{14}$\,G, which corresponds to an initial magnetic-to-fluid pressure ratio $\beta^{-1}$ of about $2\times 10^{-4}$. As before, the atmosphere floor density is set to $6.3\times10^{5}$\,g/cm$^3$ for the entire domain. 

For this test, we use a fixed spacetime metric that is not evolved in time.
The metric we use has a coordinate singularity at the origin. Nevertheless, we evolve the magnetohydrodynamic
equations in the full domain. In order to avoid singular values, 
we place the numerical grid such that the origin is located at the corners of the adjacent cells on the 
finest refinement level, and never at a cell center.
Still, the large gradients that spacetime-related variables exhibit near the 
origin pose a challenging test for the robustness of the code.

We employ fixed `box-in-box' mesh refinement, using a computational grid hierarchy 
with five cubical refinement levels. The coarsest level extends up to $528\,$km and the finest one
up to $33\,$km, with a resolution of $\sim 531.6$\,m.
For the chosen grid hierarchy, some mesh refinement boundaries are crossing the disk. This could be 
avoided in a production setting, but as test case facilitates the observation of potential artifacts.

Related to the potentially problematic mesh refinement boundaries and BH center, we test several 
evolution options. 
One is to employ the algebraic gauge condition \cite{etienne2012,etienne2010} for the vector potential, 
but without using any artificial numerical dissipation.
Next, we use the generalized Lorenz gauge (see Eq. (29) of \cite{Cipolletta2020}) still without 
dissipation. Finally, we use the generalized Lorenz gauge while applying Kreiss-Oliger dissipation 
\cite{Kreiss1973} to the evolved vector and scalar potentials. 
In particular, we use the \texttt{Dissipation} module of the {\tt Einstein Toolkit},  
with dissipation order $p=5$ and 
dissipation strength parameter $\epsilon_\text{diss} = 0.1$, which is 
a typical value used in other works (see, for e.g., \cite{Kawamura2016, Ciolfi2017, Ciolfi2019}).

For the test case at hand, the use of the algebraic gauge leads to numerical instabilities at 
the mesh refinement boundaries inside the disk. This seems related solely to the evolution, 
whereas the C2P algorithm was not causing any problems. As an additional cross-check, we verified
that the same behavior arises when using the Noble C2P scheme.
The evolution using Lorenz gauge without dissipation shows high-frequency noise at the refinement boundaries,
as well as numerical instabilities near the origin 
\footnote{The instability near the origin is most likely caused by the particular metric used for this test. 
It is worth noting that it differs significantly from the one adopted in numerical merger simulations that are using the puncture gauge.}. 
Again, the C2P algorithm was not causing problems.

\begin{figure*}
  \begin{center}
    \includegraphics[width=1.6\columnwidth]{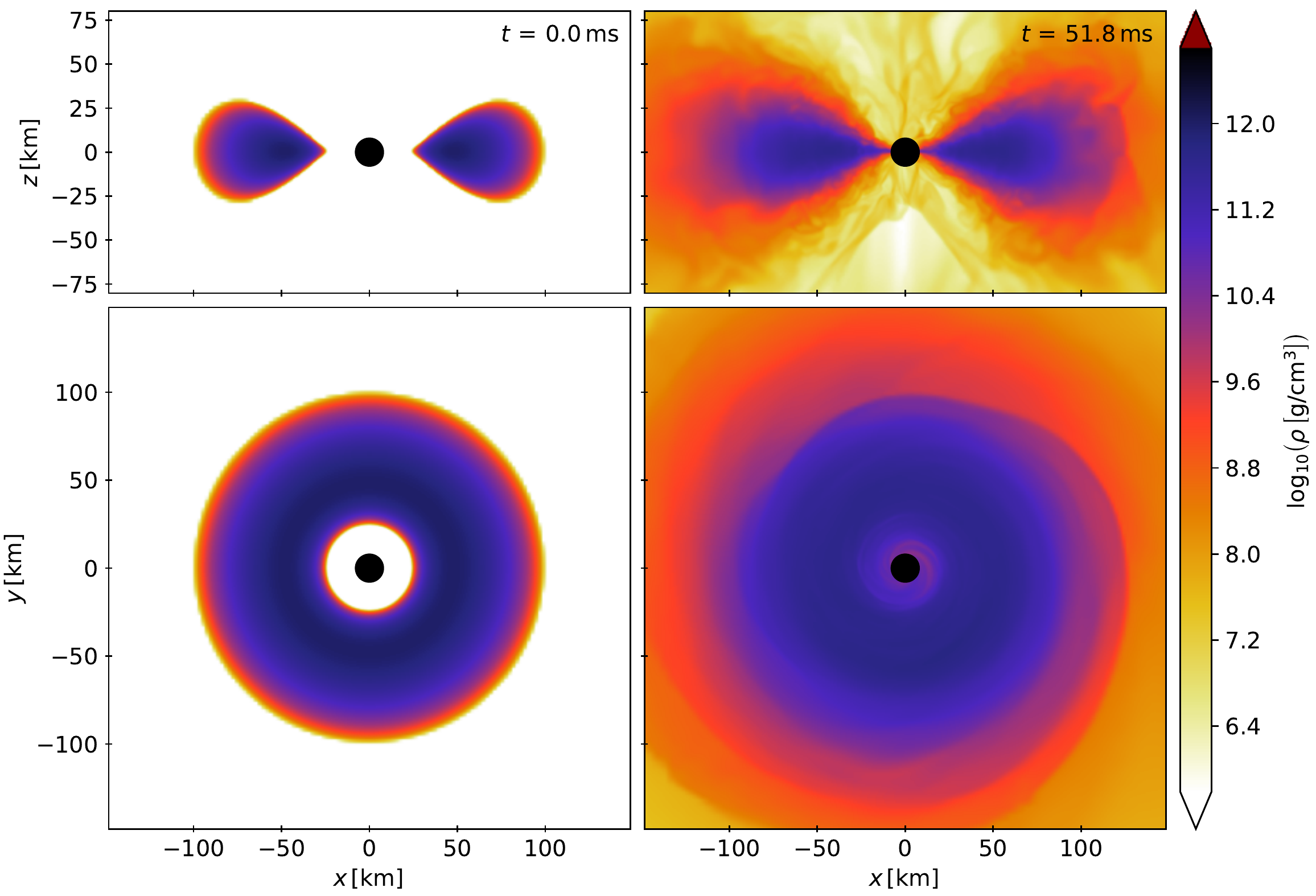}  
    \caption{Initial and final time snapshots of the rest-mass density in the meridional (top-row) and equatorial (bottom-row) planes for the magnetized Fishbone-Moncrief BH-accretion disk test. The apparent horizon is depicted in black. }
    \label{fig:fmdisk_rho2d}
  \end{center}
\end{figure*}

When using the Lorenz gauge with dissipation, we obtain a stable evolution without artifacts near the mesh
refinement boundaries.
We evolve this case for about 50\,ms. The rest-mass density distribution at initial and final time is shown in \Fref{fig:fmdisk_rho2d}. 
Within the first few ms, disk matter starts to fall into the BH, in part channeled along the BH spin axis, while the inner edge of the disk moves closer and closer to the horizon on the equatorial plane. At the same time, the outer layers of the disk begin to expand and the initial sharp outer edge of the disk is lost in favor of a layer of low density material gradually spreading up to larger and larger distances. At the end of the simulation ($\sim\!50$\,ms), after the initial transient phase, it looks as if the system is
starting to adjust towards a quasi-stationary configuration. The latter condition would however be fully achieved only via longer simulations, as reported in the literature (see, e.g., \cite{porth2019event}). 
\begin{figure*}
  \begin{center}
    \includegraphics[width=1.6\columnwidth]{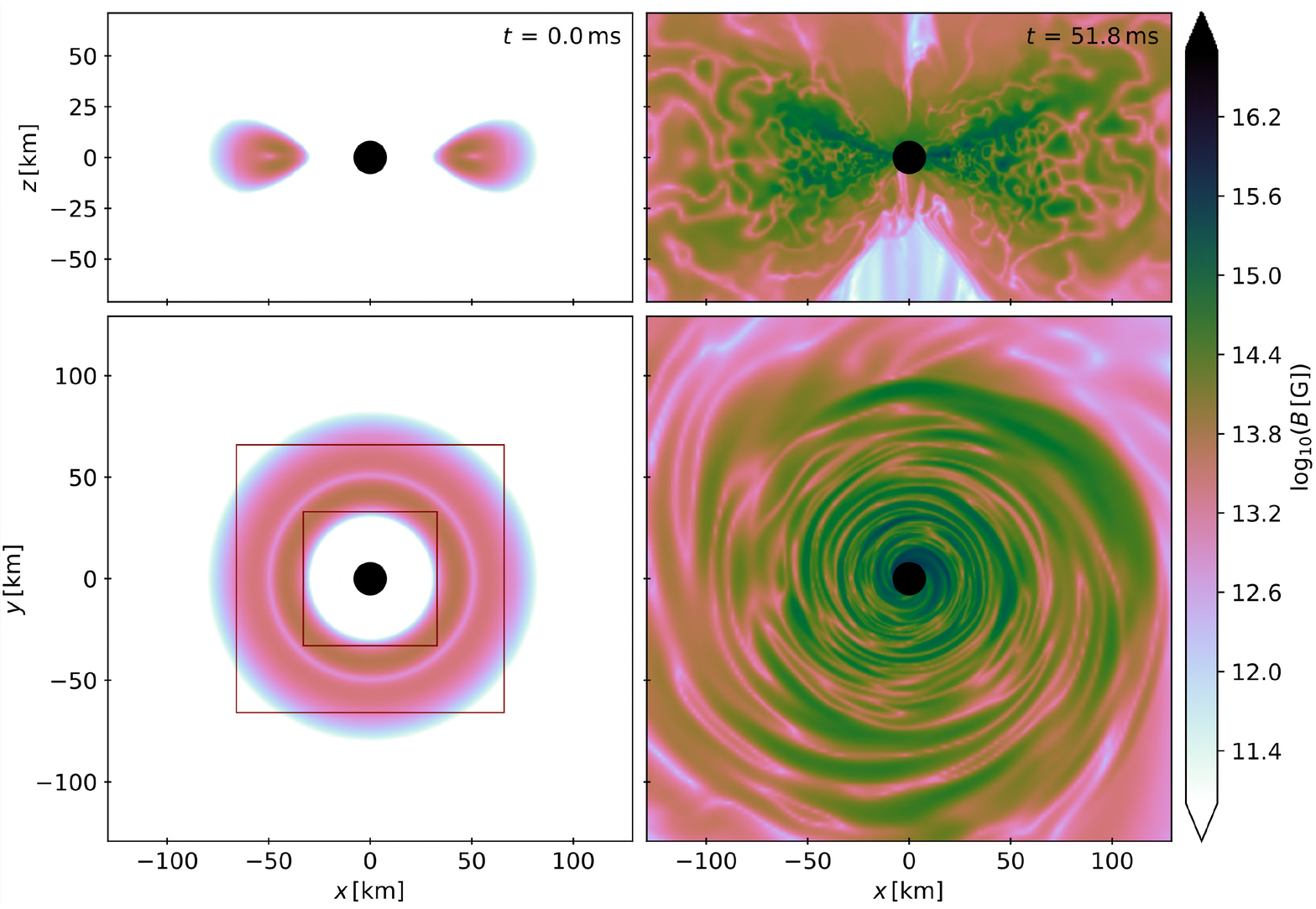}  
    \caption{Initial and final time snapshots of the  magnetic field strength within the meridional (top-row) and equatorial (bottom-row) planes, for the magnetized Fishbone-Moncrief BH-accretion disk test. The apparent horizon is depicted in black.
    In the lower left panel, we also highlight the mesh-refinement boundaries (red).}
    \label{fig:fmdisk_bfield2d}
  \end{center}
\end{figure*}

\begin{figure}
  \begin{center}
    \includegraphics[width=0.95\columnwidth]{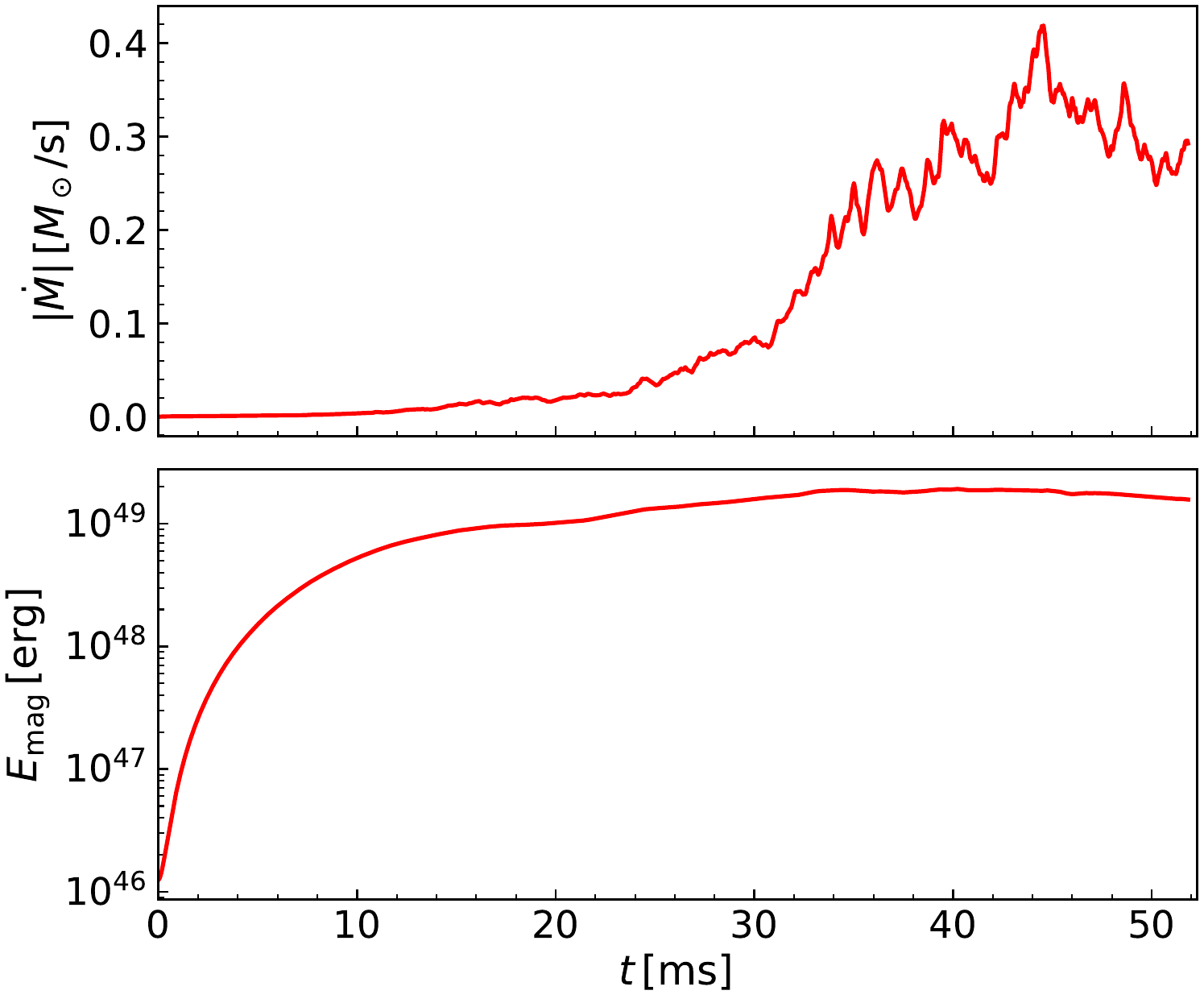} 
    \caption{Time evolution of the mass accretion rate (top-panel) and total magnetic energy (bottom-panel) for the magnetized Fishbone-Moncrief BH-accretion disk test.}
    \label{fig:fmdisk_plot1d}
  \end{center}
\end{figure}

\Fref{fig:fmdisk_bfield2d} illustrates the magnetic field strength at the initial and final time frames. For comparison, we draw the locations of the mesh refinement boundaries in one panel, illustrating that there are no significant artifacts present until the end of simulation. 
During the evolution, magnetic fields get dragged along with the fluid into the BH horizon as shown in the right-panels of \Fref{fig:fmdisk_bfield2d}.
Moreover, the magnetic field within the disk undergoes a strong amplification, particularly in the toroidal component, induced by the differential rotation and shear in the disk. The maximum magnetic field strength grows by about two orders of magnitude in 50\,ms, aided by magnetic winding, and potentially the development of the magneto-rotational instability (MRI). 
Using $\lambda_{\rm MRI} \approx (2\pi/\Omega) \times B/\sqrt{4\pi \rho}$ as an estimate for the wavelength of the fastest growing mode, we find that at 50\,ms it is resolved by 10--100 grid points in most of the disk, with local growth timescales of order 1--10\,ms.
Overall, magnetic energy grows by around three orders of magnitude as shown in Fig.~\ref{fig:fmdisk_plot1d}, and $\beta^{-1}$ increases more than the maximum field strength, reaching values up to 100. 
The reason is that the regions of low density above and below the disk become increasingly permeated by strong fields.

The top panel of \Fref{fig:fmdisk_plot1d} shows the mass accretion rate, computed from the mass flux penetrating the horizon (in detail, we consider the mass flow rate through a spherical surface of radius $\sim\!7.83$\,km that is slightly larger than the BH). 
We find that after $10$--$12$\,ms, the accretion rate starts to rapidly grow and the increase  continues up to a maximum reached around 45\,ms. Towards the end of the simulation, we observe first hints of saturation and what appears to be the beginning of a decreasing trend.
We note that on longer timescales,
a jet might form as well. The current test should not be regarded as representative for this scenario, since it would involve 
magnetizations, temperatures, and velocities that exceed those reached here.

For comparison, we repeated the test with the same setup, but employing the Noble C2P implementation 
of the Spritz code. 
We observed instabilities similarly to the results shown in Sec.~\ref{sec:dipoletov}, as well as outright C2P failures,
which occur within $1\,\milli\second$. Those problems occur in a region extending outside the horizon, 
including the inner edge of the disk. When increasing the atmosphere density to 10 (100) times the one used for
RePrimAnd test, similar failures occur within $5 (11)\,\milli\second$.

In conclusion, our RePrimAnd scheme implementation functions well in the regimes encountered within the first 50\,ms of evolution of the chosen magnetized Fishbone-Moncrief BH-accretion disk setup, for a floor density as low as $6.3\times10^{5}$\,g/cm$^3$ and a (rather low) finest resolution of $\sim\!530$\,m. It is more reliable than the old C2P implementation, which failed such a test.  
The RePrimAnd scheme is also functioning robustly in the presence of numerical instabilities in the evolution itself, caused by other 
effects such as coordinate singularities. 
The physical setup in this test is representative for science runs involving long-term evolution of post-merger accretion 
disks around BHs before (or without) jet formation.

%=====================================================================
% Conclusions
%=====================================================================
\section{Conclusions and Outlook}
\label{sec:conclusions}

In this work, we implement our new primitive variable recovery scheme RePrimAnd \cite{Kastaun2021} in the GRMHD code {\tt Spritz}. We perform a number of demanding 3D GRMHD tests representative for astrophysical scenarios, including binary neutron star mergers and BH
formation. In addition, we carried out various 1D shock tube tests with known solution up to high magnetization.
Those tests validate the correctness of the implementation as well as the stability of the interplay of the evolution scheme with the recovery scheme. We also discuss important technical details regarding the treatment of invalid states, validity
bounds of the EOS, and the evolution inside BHs without use of excision.

When evolving a TOV configuration with magnetic field confined entirely inside the star, the field evolution is well-behaved and our resolution study shows equivalent results for test cases using the RePrimAnd scheme and the commonly used scheme of Noble et al.~\cite{Noble2006}, establishing an important consistency check. 

Next, we consider the same TOV configuration, but with an initial dipolar magnetic field extending also outside the NS surface. This test shows the ability of the RePrimAnd scheme to handle  magnetized, low density environments with magnetic-to-fluid pressure ratios reaching $\sim\!10^4$. In particular, we find that the RePrimAnd scheme performs well for the extremely low artificial atmosphere density of $6.3\times10^{5}$\,g/cm$^3$ used in this test. In contrast, we find that the Noble scheme fails unless the floor density is increased by at least two orders of magnitude. 

In another test, we start from a non-magnetized uniformly rotating NS and add the same extended dipolar field after 10\,ms, then evolve the system for another 10\,ms. At the time the magnetic field is added, the outermost layers of the star have expanded and are no longer uniformly rotating. Such a differential rotation winds up the field, generating a growing toroidal component, while at larger distances the dipolar field structure remains essentially unaffected. 
The RePrimAnd scheme shows again efficient handling of magnetic-to-fluid pressure ratios as high as $\sim\!10^4$. Additionally, this test case ascertains correct behavior when employing mesh-refinement.

The crucial test case of a magnetized differentially rotating NS collapsing to a rotating BH successfully proves the robust performance of the scheme also in highly dynamical conditions and for highly magnetized environments around BHs. Moreover, it shows that the scheme is not causing problems when evolving the interior of the BH without excision. This test is fundamental for NS-NS/NS-BH merger simulations that involve a BH merger remnant surrounded by a magnetized disk, and in particular for studies related to development of magnetically driven relativistic jets by such BH-disk systems.

As our final 3D test, we simulate an accreting BH-disk system, starting from a Fishbone-Moncrief equilibrium solution where a poloidal magnetic field was added inside the disk. The system is evolved for 50\,ms, long enough to show the first signs of saturation in the growth of the accretion rate. We also observe strong magnetic field amplification, most likely caused by a mixture of magnetic winding and the development of the magneto-rotational instability.
As a side-effect, this setup also investigates the interplay between numerical evolution scheme and mesh-refinement with
refinement boundaries intersecting the disk. We find that artifacts can be avoided by using the generalized Lorenz gauge 
together with Kreiss-Oliger dissipation. 
As in the previous test, we evolve the BH interior without excision. Instead, we exploit the flexibility offered 
be the RePrimAnd scheme regarding the handling of extreme or invalid states, by employing different 
policies inside and outside the BH. 
Also in this test, the new primitive recovery scheme is not causing any problems. 

Building on the proven robustness of the RePrimAnd scheme, we are currently working on providing support for using the new scheme in {\tt Spritz} also with a fully-tabulated, composition and temperature dependent EOS. In the near future, we plan to use this additional upgrade of the scheme to perform magnetized BNS or NS-BH merger simulations as well as long-term BH-disk simulations with microphysical EOS and including an approximate treatment of neutrino radiation.

%=====================================================================
% Acknowledgements
%=====================================================================
\acknowledgments
\noindent  J. V. K. kindly acknowledges the CARIPARO Foundation for funding his Ph.D. fellowship within the Ph.D. school in Physics at the University of Padova. Most of the numerical simulations were performed on the clusters GALILEO and MARCONI at CINECA (Bologna, Italy). We acknowledge the CINECA awards under ISCRA and MoU INAF-CINECA initiatives, for the access to high performance computing resources and support (via grants {\tt IsB18\_BlueKN, IsB21\_SPRITZ, INA20\_C6A49, INA20\_C7A58}). Part of the numerical computations have been made possible through a CINECA-INFN agreement, by means of the allocations {\tt INF21\_teongrav} and {\tt INF21\_virgo}. FC acknowledges the CCRG of RIT for providing resources on the Green Prairies cluster for preliminary tests of the FM disk simulation, and also acknowledges the NASA TCAN 80NSSC18K1488 Grant for providing access to the Frontera cluster for more advanced FM disk tests.

%=====================================================================
%\bibliographystyle{apsrev4-1-noeprint}
%\bibliographystyle{iopart-num} %%%

\appendix

\section{Shock tube tests} \label{STtests}

In the following, we describe the results of basic 1D shock tube tests. Since the exact solution is known for those tests,
they are ideal to validate the basic correctness of the full evolution code. Our focus is on the robustness of the 
recovery schemes as used in the evolution code.
For a general discussion of the accuracy of the numerical evolution scheme in shock tube problems, we refer to 
\cite{Cipolletta2020} and the references therein. Here, we compare results obtained both using the Noble recovery
scheme and the new RePrimAnd scheme.

First, we repeat a series of tests known as Balsara problems 1--5 \cite{Balsara2001}, 
which have already been carried out with older Spritz code versions in \cite{Cipolletta2020}.
These tests use an ideal gas EOS.
We set the artificial atmosphere to much lower values than encountered in the simulations.
When using the RePrimAnd recovery, we also effectively disable the optional corrections (using a 
low value for $\rho_{\rm strict}$). Balsara tests 1 to 4 are evolved until a time $t = 0.4$ and Balsara 5 up to $t = 0.55$ using 1600 grid points in the domain extending from -0.5 to 0.5 along x-direction.

We find no significant differences between results with Noble and RePrimAnd schemes and neither scheme develops problems. 
The solution matches the exact solution (obtained via the code presented in \cite{Giacomazzo2006}) as well 
as expected for such tests (see \cite{Cipolletta2020}).  
We note that the Balsara tests only reach a magnetization of $\beta^{-1} {\approx} 500$.

To test the code in a more demanding regime with higher magnetization, we created two new shock tube configurations
labeled A and B in the following. The parameters are reported in Table~\ref{tabST}.
Both use an ideal gas EOS with $\Gamma=5/3$ and are evolved up to time $t = 0.4$ using 1600 grid points with the domain extension $[-0.5,0.5]$ along x-direction.
Configuration A reaches a magnetization $\beta^{-1} = 1.6 \times 10^{4}$, and configuration B 
an even higher value $10^5$.

\begin{figure*}
	\begin{center}
		\includegraphics[width=2.0\columnwidth]{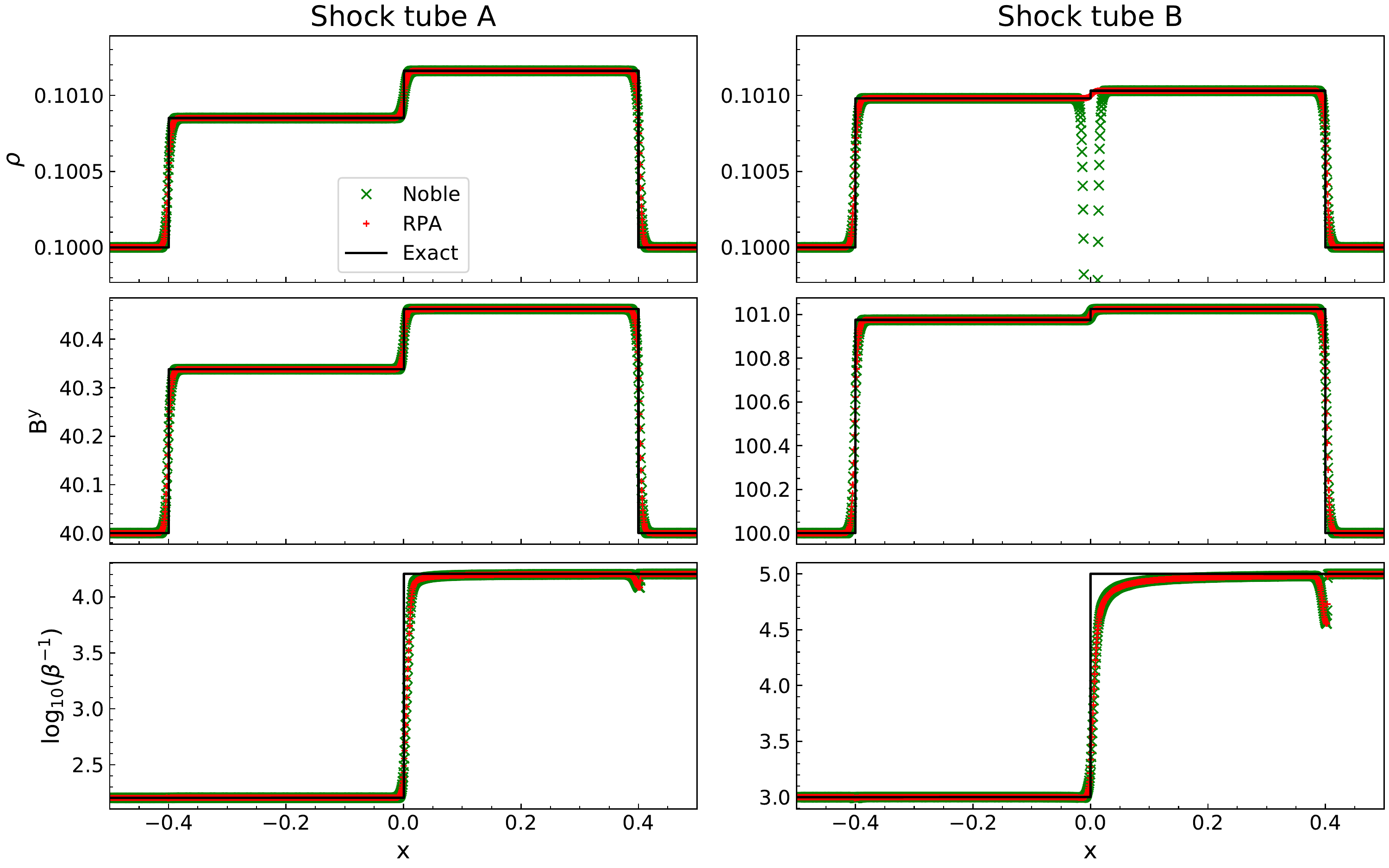} 
		\caption{Shock profiles at time $t=0.4$ for test cases A and B from Table~\ref{tabST}, 
		evolved while using the RePrimAnd and Noble C2P schemes, 
		in comparison to the exact solution computed using the code of \cite{Giacomazzo2006}.}
		\label{fig:st_plot1d}
	\end{center}
\end{figure*}

The results for the two recovery schemes are shown in Figure~\ref{fig:st_plot1d}. For test A, the plots
show exact matches between the two schemes, and typical deviations from the exact solution.
In contrast, we find significant differences between the schemes for test B. The reason is that 
the Noble scheme occasionally fails to converge and then falls back to the use of a polytropic EOS
instead of the correct ideal gas EOS. In some instances, even the fallback approximation fails to converge,
and non-converged results are used. The RePrimAnd scheme did not experience any problems with the stronger 
magnetization of this test. Correspondingly, the solution is closer to the exact one in the affected regime.

\begin{table}[!ht]
	\footnotesize
	\begin{center}
		\begin{tabular}{@{}ccccc}
			\hline	
			Test: & \multicolumn{2}{c}{ \texttt{A} } & \multicolumn{2}{c}{ \texttt{B} }  \\
			\hline
			& L & R & L & R   \\
			$\rho$ & 0.1 & 0.1 & 0.1   & 0.1   \\
			$p$     & 10.0 & 0.1     & 10.0 & 0.1  \\
			$v^x$  & 0.01 & -0.01     & 0.01   & -0.01 \\
			$v^y$  & 0.0 & 0.0     & 0.0   & 0.0  \\
			$v^z$  & 0.0 & 0.0     & 0.0   & 0.0  \\
			$B^x$  & 0.0 & 0.0   & 0.0    & 0.0  \\
			$B^y$  & 40.0 & 40.0   & 100.0    & 100.0  \\
			$B^z$  & 40.0 & 40.0    & 100.0    & 100.0 \\
			\hline
		\end{tabular}\\
		\caption{\label{tabST}Parameters for shock tube tests A and B. 
		Columns L and R refer to left and right states, and the shock propagates along the $x$-axis. }
	\end{center}
\end{table}

\clearpage

\bibliography{refs}

%=====================================================================
%=====================================================================
\end{document}